\documentclass[12pt]{article}

\usepackage{microtype} \DisableLigatures{encoding = *, family = * }

\usepackage[pdftex]{graphicx}
\usepackage[cmex10]{amsmath}
\usepackage{array}
\usepackage{mdwmath}
\usepackage{mdwtab}
\usepackage{eqparbox}
\usepackage{url}
\usepackage{amsthm}
\usepackage{amssymb}
\usepackage{csquotes}
\usepackage{float}

\usepackage{caption}

\DeclareCaptionFont{mysize}{\fontsize{8}{9.6}\selectfont}
\usepackage{graphicx}
\usepackage{caption}
\usepackage[caption=false,font=footnotesize,labelformat=simple]{subfig}

\usepackage[numbers]{natbib}
\bibpunct{(}{)}{;}{a}{,}{,}

\newcommand{\refe}[1] {{(\ref {#1})}}
\newcommand{\m}{\mbox}
\newcommand {\der}[2] {{\frac {\m {d} {#1}} {\m{d} {#2}}}}
\newcommand{\bsq}{\begin{subequations}}
        \newcommand{\esq}{\end{subequations}}
\newcommand{\bqn}{\begin{eqnarray}}
        \newcommand{\eqn}{\end{eqnarray}}

\def\eop {{\hfill $\blacksquare$}}

\newtheorem{theorem}{Theorem}[section]

\newtheorem{lemma}[theorem]{Lemma}

\usepackage[labelformat=simple]{subfig}           
  
\newcommand\reff[1]{Fig.~\ref{#1}}

\begin{document}
\title{A Control Theoretic Approach to Simultaneously Estimate Average Value of Time and Determine Dynamic Price for High-occupancy Toll Lanes}

\author{Xuting Wang \footnote{Xuting Wang is with Department of Civil and Environmental Engineering, The Pennsylvania State University, University Park, PA, 16802, USA (email: xpw5019@psu.edu).}
  and      Wen-Long Jin \thanks{Wen-Long Jin is with Department of Civil and Environmental Engineering, California Institute for Telecommunications and Information Technology, Institute of Transportation Studies, University of California, Irvine, CA, 92697, USA (email: wjin@uci.edu).}
        and Yafeng Yin \footnote{Yafeng Yin is with Department of Civil and Environmental Engineering, University of Michigan, 2350 Hayward, 2120 GG Brown, Ann Arbor, MI 48109-2125, USA (email:yafeng@umich.edu).}
}

\maketitle

\begin{abstract}
The dynamic pricing problem of a freeway corridor with high-occupancy toll (HOT) lanes was formulated and solved based on a point queue abstraction of the traffic system \cite{yin2009}. However, existing pricing strategies cannot guarantee that the closed-loop system converges to the optimal state, in which the HOT lanes’ capacity is fully utilized but there is no queue on the HOT lanes, and a well-behaved estimation and control method is quite challenging and still elusive. 

This paper attempts to fill the gap by making three fundamental contributions: (i) to present a simpler formulation of the point queue model based on the new concept of residual capacity, (ii) to propose a simple feedback control theoretic approach to estimate the average value of time and calculate the dynamic price, and (iii) to analytically and numerically prove that the closed-loop system is stable and guaranteed to converge to the optimal state, in either Gaussian or exponential manners.
%The methodology and result are novel and relevant both theoretically and practically, and the estimator/controller as well as the analytical method can enable us for better understanding and design of effective and efficient dynamic pricing strategies for real-world systems in the future.
\end{abstract}


\section{Introduction}
\subsection{Background}

High-occupancy vehicle (HOV) lanes are those reserved for cars with a minimum of two or three occupants and other qualified vehicles. The first HOV lane was implemented on Virginia's Shirley Highway busway facility (I-395) in 1969; as of 2005, HOV lanes comprised 1,305 (directional) lane miles of freeway in California, and 950 additional lane miles had been proposed for construction \cite{jang2009safety}. They can encourage car-sharing and reduce congestion, improve the people-moving capability and reliability, and lead to more efficient usage of the available roadway infrastructure and transit system \cite{fuhs2002development}. However, some HOV lanes could be underutilized, even when the corresponding general purpose (GP) lanes on the same roads are congested. For example, for more than 700 detector stations in California during PM peak hours on 128 weekdays in 2005, the flow-rates of 81\% HOV lanes were below 1400 vphpl, and most of them had speeds over 45 mph and thus were uncongested and underutilized \cite{kwon2008}.

\begin{figure}[h]
\centering
\includegraphics[width=3.3 in]{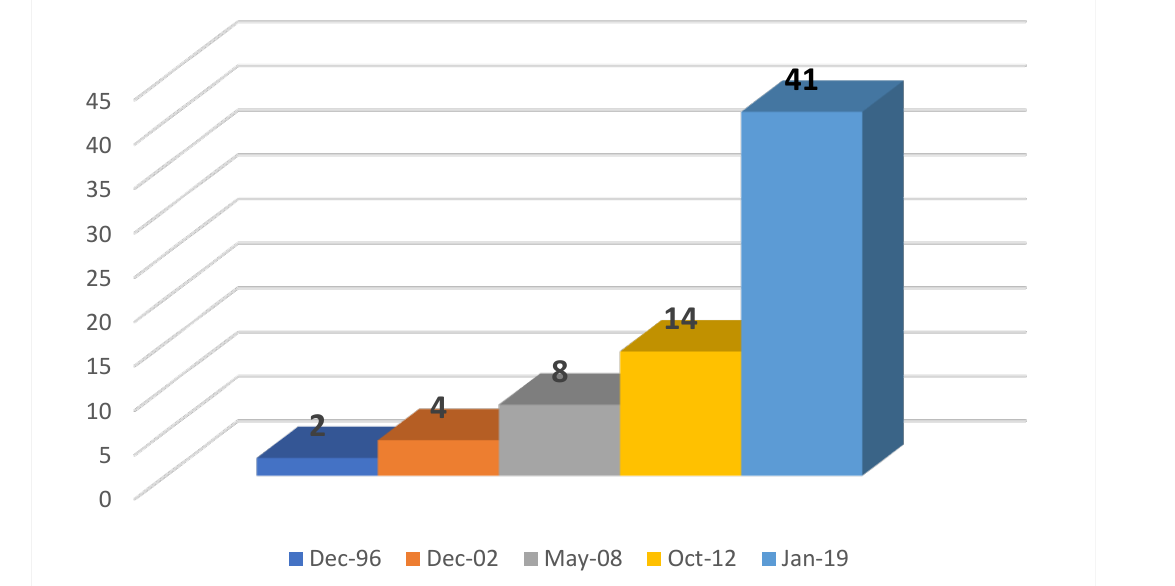}
\caption{Exponential increase in the number of HOT lanes in the U.S.}
\label{fig:HOT_dev}
\end{figure}

Congestion pricing has received more and more attention in both economics and transportation fields since the work by \cite{pigou, knight}. Many strategies (such as raising fuel prices) usually only provide short-term relieves, but only pricing strategies could manage congestion in the long run \cite{sorensen}. A type of relatively recent congestion pricing strategies are realized with high-occupancy toll (HOT) lanes, where single-occupancy vehicles (SOVs) can pay a price to use HOV lanes during peak periods. The first HOT lane has been implemented on SR-91 in California since 1995; the first HOV-to-HOT conversion project started operation on I-15 near San Diego in 1996; as of May 2012, 14 HOT lanes had been implemented, and additional 14 facilities were under construction \cite{perez2012price}; as of January 2019, 41 HOT lanes had been implemented nationwide \cite{managedlanes_2019}. As can be seen in \reff{fig:HOT_dev}, the number of HOT lanes in the U.S. has exponentially increased during the last twenty years. In addition to improving the performance of the overall system with better utilization of underutilized HOV lanes, HOT lanes can have some other benefits: (i) generating new revenue sources that can be used to support the construction of HOT lanes themselves or other initiatives and (ii) protecting environment by providing opportunities to encourage carpooling, improving transit service and moving more people in fewer vehicles at faster speeds \cite{perez2003guide}. 
Typical objectives of operating HOT lanes are: (i) maintaining a certain level of service (LOS) on the HOT lanes, which includes keeping free-flow speed, keeping zero queue, and keeping the speed above certain value; (ii) improving the overall system performance (e.g., maximizing vehicle or person throughput, which is equivalent to minimize the travel time); and (iii) maximizing revenue for the operators \cite{perez2012price}.

So far HOT lanes implemented in the real world have been operated based on heuristic pricing schemes, which are determined based on the historic and current traffic conditions. For example, the original SR-91 Express Lanes in the Orange County applied time-of-day pricing schemes to optimize the throughput of the HOT lanes at free-flow speed while generating sufficient revenue for operation and corridor improvement; since there was only one entry and one exit in each direction, the toll collection can be categorized as per use-based. Since March 2017, the Express Lanes have been extended to Riverside County, and drivers may drive the entire length or enter or exit at the county line near Green River Road; thus the prices to use the extended SR-91 Express Lanes are both time-of-day and distance-based. The prices are adjusted quarterly based on the average hourly flow-rate in the last 12 consecutive weeks \cite{SR91tollpolicy}. During Phase I of the I-15 Express lanes near San Diego, a limited number of SOVs paid a flat monthly fee to access the Express Lanes; during the ongoing Phase II (started in March 1998), the I-15 Express Lanes apply a distance-based dynamic pricing scheme for SOVs between \$0.50 and \$8.00 to maintain LOS C on the Express lanes, and the price was updated once every 6 minutes initially or 3 minutes more recently based on the level of traffic of the Express Lanes \cite{supernak2003dynamic,brownstone2003}. However, these heuristic pricing schemes cannot guarantee that the overall traffic system is at the optimal state. In some cases, they would increase the price when the HOT lanes are uncongested and severely underutilized with a very low flow-rate even if the corresponding GP lanes are congested, cannot eliminate congestion on the HOT lanes after a queue forms on them and the demand is relatively high, and may not be able to cope with accidents on the toll lanes. For example, the SR-91 Express Lanes increase the hourly toll by $\$ 1.00$ if the average flow-rate is 3300 vph or more during peak hours \cite{SR91tollpolicy}. However, the flow-rate itself cannot describe whether the traffic is congested or not. Less SOVs will pay and switch if the operators increase price when the Express Lanes are uncongested, thus the congestion on the GP lanes will be even worse.

\subsection{Literature review}

To develop more efficient pricing schemes for HOT lanes, it is critical to have a better understanding of the characteristics of such a traffic system, including how SOVs respond to the change in the price. In general, more SOVs want to pay and switch to the HOT lanes with a lower price when the overall system is congested. When the GP lanes are becoming more congested, there will be a larger travel time difference between the GP and HOT lanes, and more SOVs will choose the HOT lanes; in this case, the operators need to increase the price to keep the HOT lanes from being congested\footnote{A positive demand elasticity to the dynamic price for the HOT lanes is found in \cite{liu2011quantifying, janson2014hot}. As more SOVs enter the HOT lanes and the tolls increase, there is a positive correlation between price and usage \cite{brent2018dynamic}. A study on the MnPASS lane revealed that at lower tolls, an increase in price results in a higher HOT lane share (probability that a transponder owning SOV will use the MnPASS lane), whereas at higher tolls, an increase in price causes a decrease in HOT lane share \cite{janson2018choice}. Thus the choice model in this study is reasonable once the toll price is above a threshold value.}. So, in a corridor with HOT lanes, intuitively, SOVs' choices between the HOT and GP lanes depend on the difference in the travel times on the HOT and GP lanes, as well as congestion prices. Generally, utility (or generalized cost) is applied to connect those two factors above in discrete choice or other lane choice models. On the HOT lanes, the generalized cost of an SOV equals the product of the travel time and value of time (VOT) plus the congestion price; but on the GP lanes, it equals the product of the travel time and VOT. In economics, the VOT represents the opportunity cost of the time that a traveler spends on trips. The concept of VOT plays an important role in congestion pricing analysis as it shows user's trade-off between cost and time \cite{yang2005mathematical}. For examples, at the same price, SOVs with higher VOTs will be more willing to pay and switch to HOT lanes, and fewer SOVs will pay to use HOT lanes if the price is too high. In most cases, travelers' VOTs are unknown to the operators, and have to be estimated. Traditionally, VOTs have been estimated offline. With the maximum log-likelihood criterion, Lam and Small estimated the mean VOT based on the survey data and loop detector data on SR-91 \cite{lam2001}. Later, Brownstone et al. estimated the median VOT of \$30/hr with the revealed preference data from drivers, loop detector data and ETC data from I-15 in Southern California \cite{brownstone2003}. However, those estimation methods cannot be applied in real-time operation, since it takes time to collect data from FasTrak users. At the same time, Liu et al. found that the median VOT is dependent of the departure time during the peak hours with the data from SR-91 \cite{liu2007estimation}. Tseng and Verhoef showed that VOTs vary by time of day with the stated-preference survey data from Dutch commuters \cite{tseng2008value}. Campbell mentioned that it is difficult to  accurately quantify the true benefits of dynamic pricing projects without a quick estimation of VOTs \cite{campbell2016estimating}. Based on those observations, we conclude that it is necessary to estimate VOTs in real-time to determine effective pricing strategies.

Several types of models have been proposed to capture the lane choice behaviors subject to different travel times, VOTs, and congestion prices. Gardner et al. presented three models to calculate the proportion of SOVs switching to the HOT lanes \cite{gardner2013development}. The first model was an all-or-nothing assignment, in which all vehicles were assigned to the lane with lower generalized cost. The second model was a logit model, with the utility of each lane being the sum of travel time, congestion price and an independent and identically distributed Gumbel disturbance term. In the third model, the proportion of SOVs choosing the HOT lanes was exactly the proportion of travelers whose VOT exceeded the ratio of price and travel time difference. The last model can be regarded as an application of the user equilibrium principle, considering a Burr distribution for VOTs. Therefore, assuming that traffic conditions, including SOVs' choices of lanes and travel times on the HOT and GP lanes, can be observed, drivers' VOTs can be calculated from congestion prices, and vice versa. That is, VOTs and congestion prices can be determined simultaneously; in particular, VOTs can be estimated online, with a dynamic pricing strategy.

In literature, most studies are concerned with the first two operational objectives, which guarantee that the trip time reliability of both HOVs and paying SOVs and help to minimize the delay on the GP lanes. But they differ in their estimation of VOTs, pricing strategies as well as the underlying traffic flow and lane choice models. Yin and Lou proposed a feedback method and a self-learning method to determine the dynamic price to provide a free-flow traffic condition on the HOT lanes while maximizing the throughput of the freeway \cite{yin2009}. For both methods, they used a logit model to capture the lane choice of SOVs in a freeway segment with HOT and GP lanes, and a point queue model to capture traffic dynamics. But different from \cite{brownstone2003,lam2001}, they applied the Kalman filtering technique, an estimation method in control theory, to estimate drivers' VOTs in real time. Zhang et al. modeled the lane choice behavior of vehicles with a logit model \cite{zhang2008}. Then, they applied a piecewise feedback control model to calculate the probability for choosing HOT lanes based on different speeds on the HOT and GP lanes. And the price was estimated from the logit model. Based on the self-learning method, Lou et al. considered the impacts of lane-changing behaviors with a multi-lane hybrid traffic flow model \cite{lou2011}. Michalaka et al. formulated a robust pricing optimization problem to maximize the total throughput while controlling the congestion level on the HOT lanes  \cite{michalaka2011proactive}. The traffic dynamics were described by the cell transmission model, and the flow-rates on the GP and HOT lanes were estimated by a logit model.

However, there are some deficiencies in literature. Zhang et al. didn't consider VOTs in the lane choice model \cite{zhang2008}. VOTs are assumed known to the operators before determining the dynamic price \cite{tan2018hybrid,dorogush2015modeling}. The feedback method in \cite{yin2009} can be regarded as an application of the ALINEA strategy \cite{papageorgiou1991alinea}, which is widely used in ramp metering. But when the overall traffic is congested, a single Integral controller is not able to keep zero queue on the HOT lanes. Examples can be found in Fig. 4, 5 and 8 in \cite{yin2009}. In the self-learning method, a substantial residual queue exists on the HOT lanes in the simulation results for the self-learning method. At the same time, a theoretic study on the stability of the HOT lanes' operation is not available. In summary, existing pricing strategies cannot guarantee that the closed-loop system converges to the optimal state, in which the HOT lanes’ capacity is fully utilized but there is no queue on the HOT lanes, and a well-behaved estimation and control method is quite challenging and still elusive. 

Note that the simultaneous estimation and control problems are relatively new in the transportation field but have been studied in many other areas. In the field of economics, Taylor applied a Kalman filter to estimate the firm's long-term demand for inventoried goods in the existence of distributed lags \cite{taylor1970existence}. After estimating the demand, a feedback decision rule was applied to minimize the expected cost. An estimation and identification of parameters and control variables of electric-motor-driven motion system can be found in \cite{ohnishi1994estimation}. In space science, Habib  used an extended Kalman filter to estimate the spacecraft position and velocity, and then applied a PD controller to control the system \cite{habib2013simultaneous}.

\subsection{Study objectives}

In this study, we attempt to fill the gap and solve the estimation and control problem simultaneously in the same physical traffic system as \cite{yin2009,lou2011}. In particular, we will (i) define a new variable called the residual capacity and present a simpler formulation of the point queue model; (ii) propose a simple feedback control theoretic approach to estimate the average value of time and calculate the dynamic price; and (iii) analytically and numerically prove that the closed-loop system is stable and guaranteed to converge to the optimal state. 

HOT systems can have many stochastic attributes in traffic demands, VOTs and traffic flow models. The stochasticity in traffic demands were considered in \cite{yin2009,lou2011}. The Burr-distributed VOTs was introduced to include heterogeneous drivers in \cite{gardner2013development}. In this study, we include the randomness in the lane choice model, and numerically check the impacts of stochasticity in Section \ref{robustness}. As our estimation and control method is stable and robust, the impacts of such stochasticity are limited, and this justifies the importance of the stability property of our method. 

A list of notations is provided in Table \ref{table:notation}.

\begin{center}
\begin{table}[h]
\resizebox{0.95\textwidth}{!}{%
\begin{tabular}{|l|l|l|l|}
\hline
Variables & Definitions & Variables & Definitions \\ \hline
$q_1(t)$ & Demands of HOVs & $q_2(t)$ & Demands of SOVs \\ \hline
$q_3(t)$ & Demands of paying SOVs & $C_1$ & Capacity of the HOT lanes \\ \hline
$C_2$ & Capacity of the GP lanes & $\zeta(t)$ & Residual capacity of the HOT lanes \\ \hline
$\lambda_1(t)$ & Queue size on the HOT lanes & $\lambda_2(t)$ & Queue size on the GP lanes \\ \hline
$\epsilon$ & Infinitesimal positive number & $\Delta t$ & Time step size \\ \hline
$w_1(t)$ & Queuing time on the HOT lanes & $w_2(t)$ & Queuing time on the GP lanes \\ \hline
$w(t)$ & \begin{tabular}[c]{@{}l@{}}Queuing time difference between the\\  GP and HOT lanes\end{tabular} & $g_1(t)$ & Throughput of the HOT lanes \\ \hline
$g_2(t)$ & Throughput of the GP lanes & $u(t)$ & Time-dependent price for paying SOVs \\ \hline
$\pi^\star$ & True average VOT & $\pi(t)$ & Estimated average VOT \\ \hline
$\alpha^*$ & A scale parameter in the logit model & $\eta$ & \begin{tabular}[c]{@{}l@{}}A random variable to capture\\  randomness in the logit model\end{tabular} \\ \hline
$K_1$ & Integral controller coefficient for $\lambda_1(t)$ & $K_2$ & Integral controller coefficient for $\zeta(t)$ \\ \hline

\end{tabular}%
}
\caption{List of notations}
\label{table:notation}
\end{table}
\end{center}

The rest of this paper is organized as follows. In Section \ref{system}, we define a new variable called the residual capacity of the HOT lanes. Then, we describe the system dynamics for the HOT and GP lanes based on the residual capacity. We also define the simultaneous estimation and control problem for the HOT lanes. In Section \ref{solution_control}, we provide the solution of the control problem with constant demand, assuming the operators knows the true average VOT of SOVs before making the pricing schemes. In Section \ref{yin_lou}, we provide a detailed review of two control methods in \cite{yin2009}. In Section \ref{feedback_method}, we present a control theoretic approach to estimate the average VOT of SOVs, and calculate the dynamic price with the estimated average VOT and the travel time difference. In Section \ref{properties}, we analyze the equilibrium state and stability of the closed-loop control system. In Section \ref{simulation}, we first provide numerical examples to show that the method is effective, and compare it with the methods in \cite{yin2009}. Then, we test the robustness of the control system with respect to randomness in demands and parameters or variables in the lane choice model. At the same time, we numerically prove that the closed-loop system is stable and guaranteed to converge to the optimal state. We also examine the effect of the scale parameter in the lane choice model. In Section \ref{conclusion}, we conclude the study and provide future research topics.

\section{Definitions, system dynamics, and problem statement}\label{system}

\begin{figure}[h]
\centering
\includegraphics[width=3.5in]{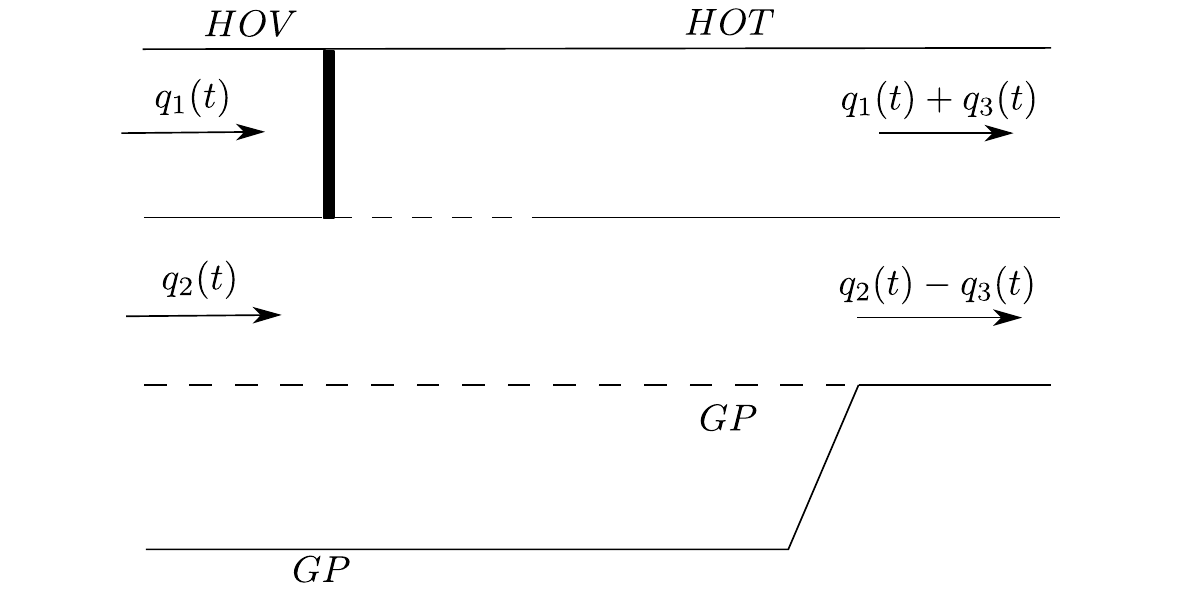}
\caption{Illustration of the traffic system with HOT lanes}
\label{fig:traffic system}
\end{figure}

We consider a freeway corridor between an origin and a destination, as shown in \reff{fig:traffic system}. The downstream is uncongested initially. The freeway has two types of lanes: the HOT and GP lanes. There is one bottleneck on the GP lanes, but not on the HOT lanes. HOVs can use the HOT lanes for free, but SOVs have to pay a price to use the HOT lanes. Thus we separate SOVs into paying SOVs that pay a price to use the HOT lanes, and non-paying SOVs that stay on the GP lanes all the time. Therefore, the HOVs and paying SOVs share the HOT lanes, but the non-paying SOVs use the GP lanes. There are two sets of loop detectors on the freeway. The first set of detectors is installed before the toll-tag reader to detect the demand on the HOV and GP lanes; while the second set is installed after the reader to detect the demand on the HOT and GP lanes. 
\vspace*{-10pt}

\subsection{Definitions of variables}
We define variables describing road and traffic characteristics as follows:
\begin{itemize}
\item Traffic demand: $q_1(t)$ and $q_2(t)$ are demands of HOVs and SOVs, respectively, at time $t$.
\item Capacities: $C_1$ and $C_2$ are the capacities of the HOT and GP lanes, respectively.
\item Queue sizes: $\lambda_1(t)$ and $\lambda_2(t)$ are the queue sizes, i.e., number of vehicles in queue, on the HOT and GP lanes, respectively.
\item Queuing times: $w_1(t)$ and $w_2(t)$ are the queuing times for vehicles leaving at $t$ on the HOT and GP lanes, respectively; $w(t)=w_2(t)-w_1(t)$ is the queuing time difference between the GP and HOT lanes.
\end{itemize}

We have the following two assumptions regarding the demand patterns. First, the total demand exceeds the total capacity during the study period $t \in [0, T]$; i.e.,
\begin{equation*}
q_1(t)+q_2(t) > C_1+C_2;
\end{equation*}
therefore, the overall traffic system is congested.
 Second, the demand of HOVs is below the capacity of the HOT lanes, i.e., 
 \begin{equation}
 q_1(t) < C_1. \label{eqn:q1}
 \end{equation}
 Under these assumptions, if SOVs are not allowed to use the HOV lanes, the HOV lanes will be underutilized and the GP lanes congested. By introducing HOT lanes, however, we can charge some SOVs to use the underutilized HOV lanes. Thus we denote the demand of paying SOVs by $q_3(t)$, and the demand on the GP and HOT lanes should be $q_2(t) - q_3(t)$ and $q_1(t) + q_3(t)$, respectively. Then, we define the residual capacity of the HOT lanes by
\begin{equation}\label{zeta}
\zeta(t)=C_1-q_1(t)-q_3(t).     
\end{equation}
We further denote the time-dependent price for paying SOVs  by $u(t)$.

\subsection{Models of system dynamics}
\subsubsection{Traffic flow model}

We apply the point queue model to model traffic dynamics \cite{vickrey1969,jin2015point}. In the point queue model, all drivers follow the first-in-first-out (FIFO) rule, and the travel time is composed of the free-flow travel time and the queuing time. For vehicles in a homogeneous freeway segment, their travel time difference only exists in the queuing time on the HOT and GP lanes. 
\vspace*{-10pt}
\begin{figure}[h]
\centering
\includegraphics[width=3.5in]{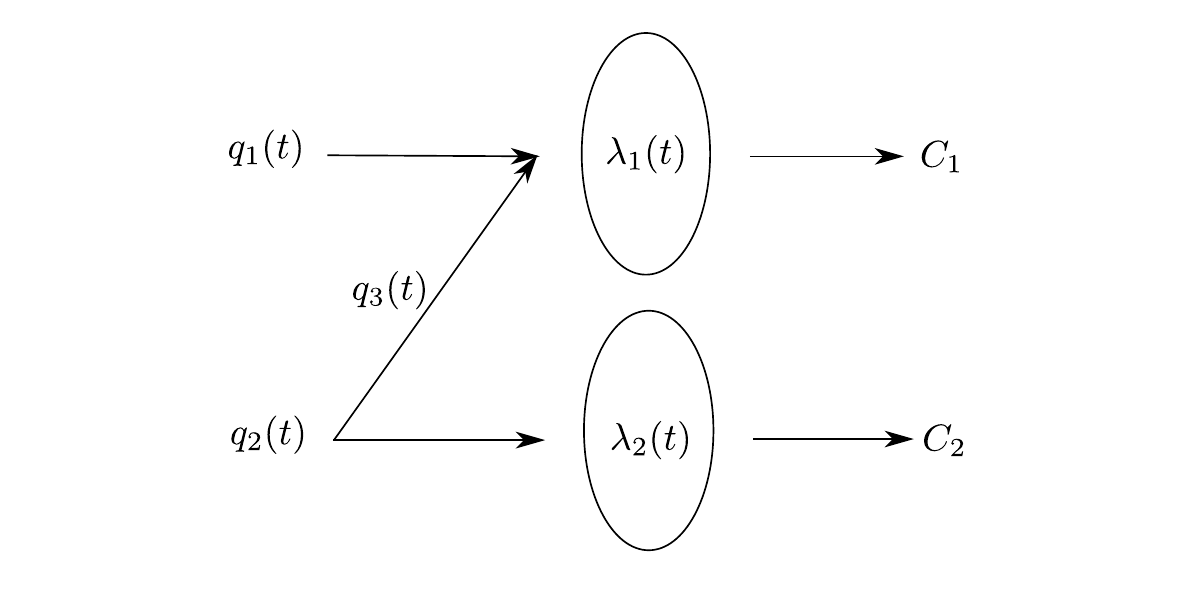}
\caption{Point queue representation of the traffic system with HOT lanes}
\label{fig:pq}
\end{figure}

For the HOT and GP lanes in this system (see \reff{fig:pq}), the dynamics of the two point queues are described by the following ordinary differential equations:
\begin{subequations}
\begin{align}
\der{}{t} \lambda_1(t) &=\max\{q_1(t)+q_3(t)-C_1, -\frac{\lambda_1 (t)}{\epsilon}\},\\
\der{}{t} \lambda_2(t) &=\max\{q_2(t)-q_3(t)-C_2, -\frac{\lambda_2 (t)}{\epsilon}\}.
\end{align}
\label{pqmodel}
\end{subequations}
With the definition of $\zeta(t)$ in \refe{zeta}, we rewrite \refe{pqmodel} as
\begin{subequations}
\begin{align}
\der{}{t} \lambda_1(t) &=\max\{-\zeta(t), -\frac{\lambda_1 (t)}{\epsilon}\}, \label{newpqmodel_1}\\
\der{}{t} \lambda_2(t) &=\max\{q_1(t)+q_2(t)-C_2-C_1+\zeta(t), -\frac{\lambda_2 (t)}{\epsilon}\},
\end{align}
\label{newpqmodel}
\end{subequations}
where $\epsilon=\lim_{\Delta t \to 0^+} \Delta t$ is an infinitesimal number and equals $\Delta t$ in the discrete form. Then, the discrete version of the point queue model becomes
\begin{subequations}
\begin{align}
\lambda_1(t+\Delta t) &=\max\{-\zeta(t) \Delta t + \lambda_1 (t) , 0\},\\
\lambda_2(t+\Delta t) &=\max\{(q_1(t)+q_2(t)-C_2-C_1+\zeta(t))\Delta t + \lambda_2 (t), 0\}.
\end{align}
\label{pqmodel-discrete}
\end{subequations}
From \refe{newpqmodel_1}, we can see that the queue changing rate on the HOT lanes is determined by either the existing queue size or the residual capacity.

In addition, the throughputs of the HOT and GP lanes are given by
\begin{subequations}\label{throughput}
\begin{align}
g_1(t)&=\min\{C_1-\zeta(t)+\frac{\lambda_1(t)}\epsilon,C_1\}, \label{HOT_throughput}\\
g_2(t)&=\min\{q_1(t)+q_2(t)-C_1+\zeta(t)+\frac{\lambda_2(t)}\epsilon,C_2\}. \label{GP_throughput}
\end{align}
\end{subequations}

So, the queuing times on the HOT and GP lanes are
\begin{subequations}
\begin{align}
w_1(t)=\frac{\lambda_1 (t)}{C_1},\\
w_2(t)=\frac{\lambda_2 (t)}{C_2}.
\end{align}
\label{waiting}
\end{subequations}
After obtaining the queuing times on two types of lanes, the queuing time difference, $w(t)$, can be calculated by
\begin{equation}\label{waiting diff}
w(t)=w_2(t)-w_1(t)=\frac{\lambda_2(t)}{C_2}-\frac{\lambda_1(t)}{C_1}.
\end{equation}

\subsubsection{Lane choice model}

For the drivers in this system, the proportion of SOVs choosing the HOT lanes, $Pr(t)$, is represented by a logit model:
\begin{equation*}
Pr(t) = \frac{\exp(\alpha^* V_{HOT}(t))}{\exp(\alpha^* V_{HOT}(t))+\exp(\alpha^* V_{GP}(t))},
\end{equation*}

where $V_{HOT}(t)$ and $V_{GP}(t)$ represent the measurable utility of drivers using the HOT and GP lanes at time step $t$; and $\alpha^*$ is a scale parameter, which determines the variation of a Gumbel distribution.  %$\alpha^*$ is the marginal effect of travel time on SOVs’ utility, $\pi^*$ is SOVs' true average VOT, and $\gamma^*$ captures other factors affecting the utility. They are properties of the SOV drivers, and are assumed to be constant in this study. 
%We assume that the drivers have full information of the travel time and dynamic price. We denote the free-flow travel time by $t_f$. 
%Then, the measurable utility for paying SOVs is $-(u(t)+ \pi^* (t_f+w_1(t)))$, and that for non-paying SOVs is $-\pi^* (t_f+w_2(t))$, where 
$\pi^*$ represents the true VOT, and it is not affected by the scale parameter \cite{train2009discrete}. We introduce a random variable, $\eta$, to capture the randomness in  the VOTs, $\pi^*$, the detection errors or imperfect information in the queuing time difference, $w(t)$, and the differences between the implemented and calculated prices, $u(t)$. In particular, $(1+\eta)\pi^*$ can be interpreted as a random or heterogeneous VOT, $(1+\eta)w(t)$ as the detected or estimated queuing time difference, and $u(t)/(1+\eta)$ as the implemented price. Then, the lane choice model for SOVs becomes
\begin{equation}
Pr(t) = \frac{1}{1+\exp(\alpha^*(u(t)-(1+\eta) \pi^* w(t)))}.
\label{choice}
\end{equation}

When the total demand of SOVs is $q_2(t)$, the demand of paying SOVs is
\begin{equation}
q_3(t) = \frac{q_2(t)}{1+\exp(\alpha^*(u(t)-\pi^* (1+\eta)w(t)))}.\label{choice-q3}
\end{equation}

With the definition of $\zeta(t)$, \refe{choice-q3} is rewritten as
\begin{equation}
\zeta(t) = C_1-q_1(t)-\frac{q_2(t)}{1+\exp(\alpha^* (u(t)-\pi^*(1+\eta) w(t)))}. \label{newchoice-q3}   
\end{equation}

\subsection{Simultaneous estimation and control problem}\label{simultaneous problem}

In this study, we only consider the first two objectives for operating the HOT lanes, which aim to optimize the whole system's performance without sacrificing the HOT lanes' LOS  \cite{perez2003guide}: (i) keeping zero queue on the HOT lanes, and (ii) maximizing the HOT lanes' throughput. Note that maximizing the HOT lanes' throughput is equivalent to maximizing the whole system's throughput in this case, since the GP lanes are always congested and their throughput is fixed.

When both of the above objectives are met at the same time, we refer to the traffic state as an optimal state. That is, in an optimal state, we have: (i) 
\begin{equation}
\lambda_1(t)=0,
\label{ob1}
\end{equation}
and (ii)
\begin{equation}
 g_1(t)=C_1.    
\label{ob2}
\end{equation}

\begin{lemma}
In the optimal state, the residual capacity on the HOT lanes is given by
\begin{equation}
\zeta(t) = 0.  \label{optimal_zeta} 
\end{equation}
\end{lemma}

{\em Proof}: Since $\lambda_1(t)=0$ in the optimal state, \refe{HOT_throughput} is equivalent to
\[
g_1(t)=\min\{C_1-\zeta(t), C_1\}.
\]
The maximum throughput of the HOT lanes occurs if and only if $\zeta (t)\leq 0$. Also, $\zeta(t)$ cannot be negative in the optimal state from \refe{newpqmodel_1}. So, at the optimal state, $\zeta (t) = 0$. \eop

Even though the optimal queue size is zero, and the optimal demand of paying SOVs can be calculated from \refe{zeta} and \refe{optimal_zeta}, the operators cannot simply force SOVs to switch into or out of the HOT lanes to achieve these objectives. As in \cite{yin2009}, we are interested in finding an appropriate pricing scheme that can drive the system to the optimal state, since the price influence SOVs' lane choice behavior according to \refe{choice-q3}, which in turn would influence the HOT lanes' queue size according to \refe{pqmodel}. In this sense, the congestion price is an actuation signal to the system, and this is a control problem \cite{astrom}. In addition, we are interested in estimating the average VOT of SOVs simultaneously since VOT is a key parameter in the lane choice model.

\section{Solution of the control problem with constant demand and VOT}\label{solution_control}
We consider a simple case when the operators know the true average VOT of SOVs, then the problem is simplified as a control problem. We assume the demand of HOVs and SOVs are time-independent. The demand of HOVs is constant at $q_1 < C_1$, and the demand of SOVs is constant at $q_2>C_2+C_1-q_1$. These are consistent with the two assumptions regarding the demand patterns in Section \ref{system}. In this case, if we do not allow SOVs to use the HOT lanes, then the GP lanes are congested, but the HOT lanes are underutilized.

With appropriate pricing schemes, the system reaches the optimal state when the two operational objectives stated in Section \ref{simultaneous problem} are met at the same time. That is, (i) 
\begin{equation*}
\lambda_1(t)=0;
\end{equation*}
i.e. there is no queue on the HOT lanes; and
(ii)
\begin{equation*}
\zeta(t)=0;
\end{equation*}
i.e., there is no residual capacity on the HOT lanes. 

Initially both lanes are uncongested, i.e., $\lambda_1(0)=\lambda_2(0)=0$. For the HOT lanes, the demand equals the capacity. According to \refe{newpqmodel_1}, the queue changing rate is zero, so there will be no queue on the HOT lanes. Then, the queuing time on the HOT lanes  $w_1(t)=0$. At the same time, the demand on the GP lanes is $q_2-(C_1-q_1)$ in the optimal state. Based on \refe{pqmodel-discrete} and \refe{waiting}, the queue size on the GP lanes is $\lambda_2(t)= (q_2+q_1-C_1-C_2)t$, and the queuing time on the GP lanes is $w_2(t)=\frac{q_1+q_2-C_1-C_2}{C_2} t$. Then $w(t)=w_2(t)=\frac{q_1+q_2-C_1-C_2}{C_2} t$. According to \refe{newchoice-q3}, the lane choice model is written as
\begin{equation*}
C_1-q_1-\frac{q_2}{1+\exp(\alpha^* (u(t)-\pi^* \frac{q_1+q_2-C_1-C_2}{C_2}t))}=0,
\end{equation*}
which leads to
\begin{equation}
u(t)=\frac{q_1+q_2-C_1-C_2}{C_2} \pi^* t+\frac{\ln\frac{q_1+q_2-C_1}{C_1-q_1}}{\alpha^*}.
\label{price_constant}
\end{equation}

This analytical result shows that the price should increase linearly with constant demand pattern and known average VOT. In this paper, we assume that the operators know the SOVs follow a logit model for the lane choice, but they do not know that the true average VOT. In this sense, we need to estimate the average VOT to determine the appropriate pricing scheme for the HOT lanes. We denote $\pi(t)$ as the estimated average VOT. Then, the operators should replace $\pi^*$ by $\pi(t)$ in \refe{price_constant} to calculate the price.

\section{Review of Yin and Lou's methods}\label{yin_lou}

This section provides a review of \cite{yin2009}. In this paper, they proposed two dynamic pricing strategies for operating HOT lanes to provide an uncongested traffic condition on the HOT lanes while maximizing the freeway's throughput.

\subsection{Feedback method}

In the feedback method, one loop detector was required downstream of the toll reader to detect the occupancy on the HOT lanes. The dynamic price on HOT lanes, $u(t)$, was calculated by an I-controller: 
\begin{equation*}
u(t+\Delta t)=u(t)+ K_I (O_{HOT}(t)-O_{HOT}^*(t)),
\end{equation*}
where $O_{HOT}(t)$ was the measured occupancy on the HOT lanes at time step $t$, $O_{HOT}^*(t)$ was the desired occupancy, and was usually equal or slightly less than the critical occupancy. The error term was the difference between the desired occupancy and the measured occupancy. $K_I$ was the coefficient for the integral controller (I-controller). The framework for operating HOT lanes could be described as \reff{fig:feedbacky}. Due to the application of the point queue model, the information of occupancy is not available, so they used the demand of the HOT lanes as the state variable in the simulation. The error term was replaced by the difference between the actual and desired demand on the HOT lanes. Then, the control logic becomes
\begin{equation}
u(t+\Delta t)=u(t)+ K_I (q_{HOT}(t)-q_{HOT}^*(t)).  
\label{feedback}
\end{equation}
When $q_{HOT}(t) > q_{HOT}^*(t)$, the price would increase; when $q_{HOT}(t) < q_{HOT}^*(t)$, the price would decrease; and when $q_{HOT}(t) = q_{HOT}^*(t)$, the price would be constant. 

Even though the feedback method is simple to implement, it fails to achieve the two objectives by directly calculating the price. Let's consider the same demand pattern in Section \ref{solution_control}. When the two objectives are met at the same time, $q_{HOT}(t) = q_{HOT}^*(t)$, then the price would be constant according to \refe{feedback}. According to \refe{choice-q3}, more SOVs are willing to pay and switch, thus the HOT lanes would be congested. So, the control system is unstable.

\subsection{Self-learning method}

\begin{figure}[h]
  \centering
  \subfloat[The feedback method]{
    \includegraphics[width=0.54\textwidth]{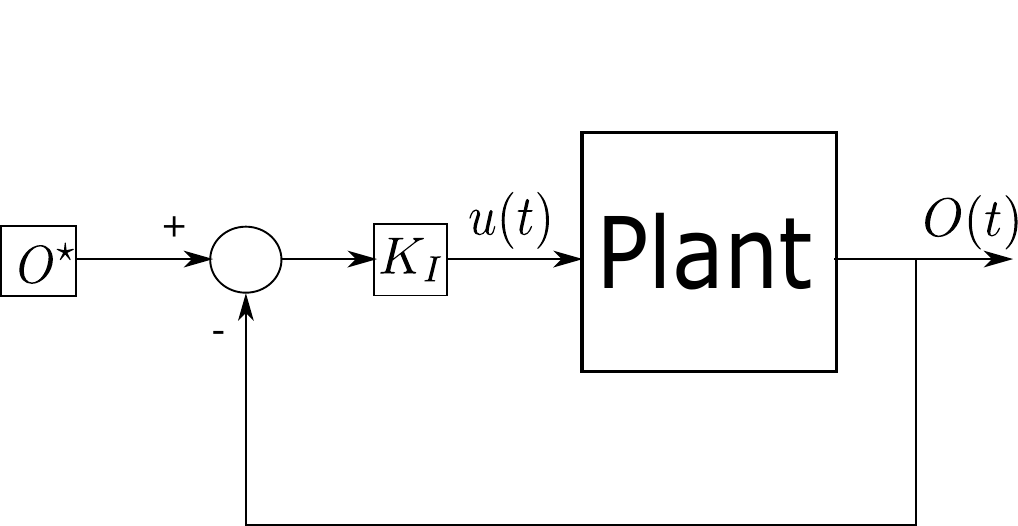}
    \label{fig:feedbacky}
  }
  \hfill
  \subfloat[The self-learning method]{
    \includegraphics[width=0.42\textwidth]{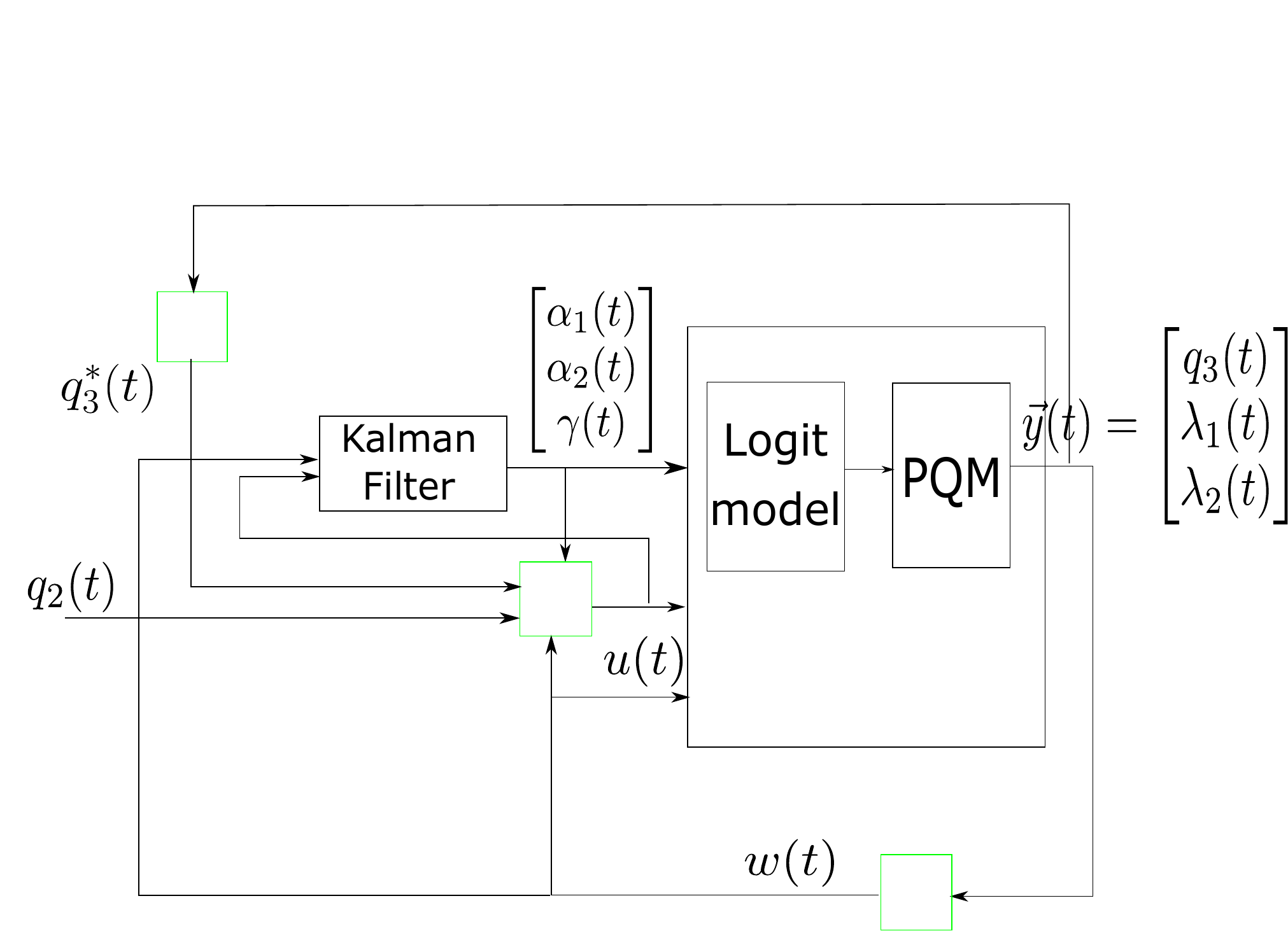}
    \label{fig:self}
  }
  \caption{Block diagrams of two methods in \cite{yin2009}}
  \end{figure}

In the self-learning method, two sets of loop detectors were required. The first set of detectors was installed before the toll-tag reader to detect the demand on the HOV and GP lanes; while the second set of detectors was installed after the reader to detect the demand on the HOT and GP lanes. They applied a logit model to describe SOV's lane choice and point queue models to capture traffic dynamics. In each step, they first learned SOV drivers' willingness to pay (WTP) by mining the loop detector data, and then determined the price based on the demand, estimated travel times, and the calibrated WTP. When calibrating the WTP, they applied a discrete Kalman filter to estimate the parameters in the logit model. And the framework is shown in \reff{fig:self}.

$\alpha_1$ and $\alpha_2$ represented the marginal effect of travel time and price on drivers' utility respectively, and $\gamma$ captured other factors that affect the WTP. $\alpha_1/\alpha_2$ was the trade-off between time savings and prices, i.e., VOT. In each step, they updated the Kalman gain and error covariance matrix in the Kalman filter. Then, they determined the dynamic price for the HOT lanes based on the logit model:
\begin{equation}
u(t)=\frac{\ln\frac{q_2(t)-q_{HOT}^*(t)}{q_{HOT}^*(t)}+\alpha_1(t)w(t)-\gamma(t)}{\alpha_2(t)},
\label{self}
\end{equation}
where $q_{HOT}^*(t)$ was the optimal demand on the HOT lanes, and $w(t)$ was the travel time difference on the GP and HOT lanes at time $t$.

When we consider the same constant demand as Section \ref{solution_control}. In the optimal state, $q_{HOT} = C_1$. Then, the queue size on the GP lanes is $\lambda_2(t)= (q_2+q_1-C_1-C_2)t$, and the travel time difference is $w(t)=\frac{q_1+q_2-C_1-C_2}{C_2} t$. Based on \refe{self}, the price should be
\begin{equation}
u(t)=\frac{\ln\frac{q_2-C_1}{C_1}+\alpha_1\frac{q_1+q_2-C_1-C_2}{C_2}t-\gamma}{\alpha_2}. 
\label{sta2}
\end{equation}
Since the price is calculated from the logit model, and it only influences the lane choice for SOVs, $C_1-q_1$ should replace $C_1$ in the log term in \refe{sta2}. 

Note that there is an inconsistency in the simulation setup in \cite{yin2009}. In both scenario 1 and 2, the average demand of HOVs is 300 vph for the feedback method, but 600 vph for the self-learning method. This can be seen by comparing Fig. 4a and 7a for scenario 1 and Fig. 5a and 9a for scenario 2. Such an inconsistency can also be confirmed by comparing the corresponding queue sizes.

\section{A new control theoretic approach}\label{feedback_method}

In this study, we are interested in simultaneously estimating the average VOT of SOVs and calculating the dynamic price for the HOT lanes to achieve those two operation objectives. We want to combine the advantages of the two methods in \cite{yin2009}. Similar to the self-learning method, we first estimate the average VOT, and then calculate the dynamic price. However, we want to take advantage of the simplicity of the feedback controller, and apply it in the estimation process. Basically, given the accurate estimation in travel time, if we overestimate the VOT, less SOVs would choose the HOT lanes; and if we underestimate the VOT, more SOVs would switch to the HOT lanes. This gives us the guideline for designing the controller: we should increase the estimated VOT when the HOT lanes are congested, and decrease it when the HOT lanes are underutilized.

\begin{figure}[htbp]
  \centering
  \subfloat[Block diagram of the control system]{
    \includegraphics[width=0.54\textwidth]{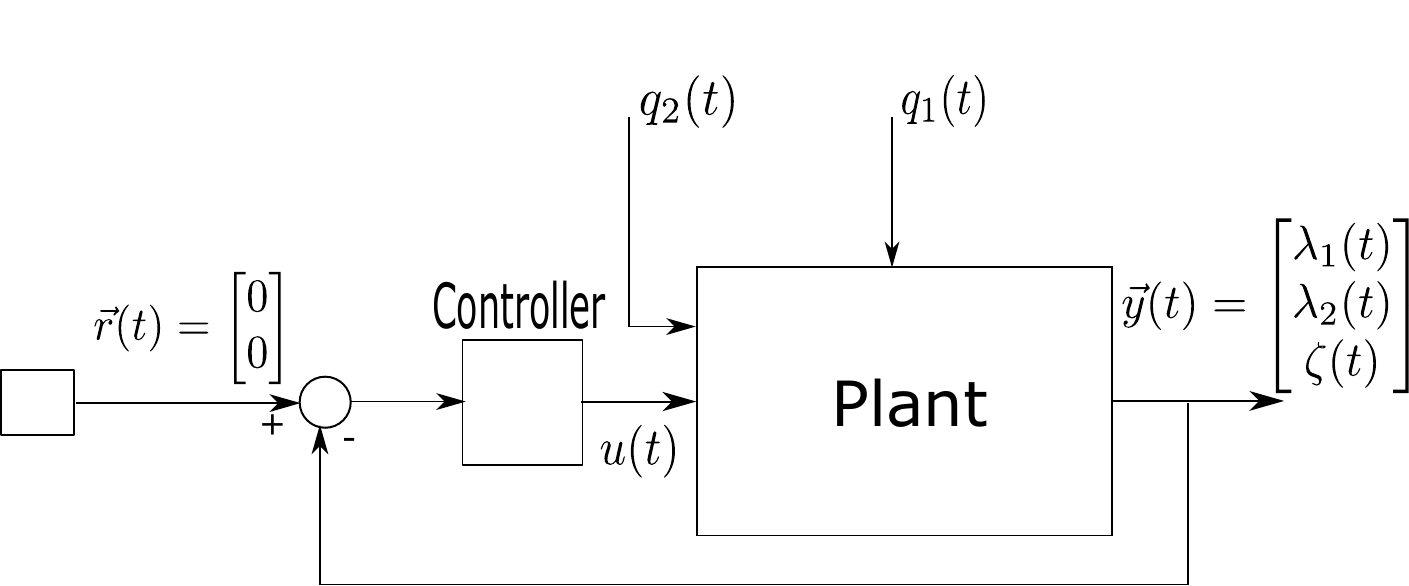}
    \label{fig:PQMdiagram}
  }
  \hfill
  \subfloat[Block diagram of the plant]{
    \includegraphics[width=0.42\textwidth]{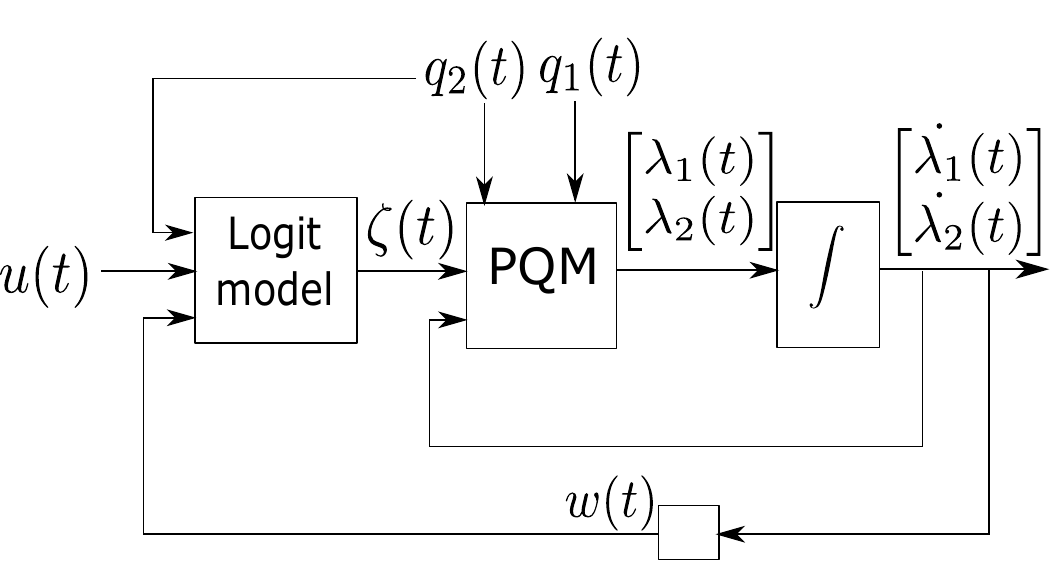}
    \label{fig:plant1}
  }
  \caption{Block diagram of the control system}
  \end{figure}

The block diagram of the control system is shown in \reff{fig:PQMdiagram}. This is a feedback system, in which the price and the system dynamics are connected together such that each system influences the other and their dynamics are strongly coupled \cite{astrom}. Let's recall the objectives for operating the HOT lanes (\ref{ob1} and \ref{optimal_zeta}): $\lambda_1(t)=0$ and $\zeta(t)=0$. They are the reference signals of the system, $\vec{r}(t)$. Then, $\lambda_1(t)$ and $\zeta(t)$ are the two error signals related to the traffic condition on the HOT lanes. With appropriate pricing schemes, $u(t)$, we can achieve those objectives. The plant has a logit model and a point queue model, and the detailed block diagram is shown in \reff{fig:plant1}. Inside the plant, the inputs for the lane choice model are $u(t)$, $q_2(t)$ and $w(t)$, and the output is $\zeta(t)$; for the traffic model, the inputs are $q_1(t)$, $q_2(t)$ and $\zeta(t)$, and the outputs are $\lambda_1(t)$, $\lambda_2(t)$ and $w(t)$.

\subsection{A feedback estimation method}\label{newcontroller}

PID control is the most common way of using feedback in engineering systems so far. In traffic operation, PID control is widely used in ramp metering \cite{papageorgiou1991alinea} and variable speed limits \cite{zhang2006simple}. In this paper, we want to apply it to solve the feedback-based estimation problem. Since $\pi^*$ is a constant value, we apply an I-controller to estimate it. The operators can gradually estimate $\pi^*$ of SOVs through learning their reaction to the dynamic price and traffic condition. 

We assume that the operators have full information of the flow-rate and queue size on the HOT lanes. For time-dependent demand, the I-controller is designed as
\begin{equation}\label{estimator2}
\der{}{t} \pi(t)=K_1 \lambda_1(t) - K_2 \zeta(t).
\end{equation}

If there is a queue on the HOT lanes, i.e., $\lambda_1(t)>0$, we should increase the estimated average VOT of SOVs. If the HOT lanes are underutilized, i.e., $\zeta(t) > 0$, we should decrease the estimated average VOT. Therefore, all the coefficients in \refe{estimator2}, including $K_1$ and $K_2$, should be positive. In the optimal state, the error signals are zero, so $\pi(t)$ would be constant.
\vspace*{-10pt}

\subsection{Calculation of the dynamic price}

With the estimated parameters in the previous subsection, we calculate the dynamic price for the HOT lanes as in \refe{price_constant}, except that the true parameters are unknown and replaced by the estimated values, and the travel time difference is replaced by $w(t)$: 
\begin{equation}\label{price}
u(t)=\pi(t)w(t)+\frac{\ln\frac{q_1(t)+q_2(t)-C_1}{C_1-q_1(t)}}{\alpha^*}.
\end{equation}
Since $\pi(t) \neq \pi^*$ initially, we would expect some fluctuation in the traffic condition and the dynamic price even if the demand is constant.

\begin{figure}[h]
\centering
\includegraphics[width=3.5in]{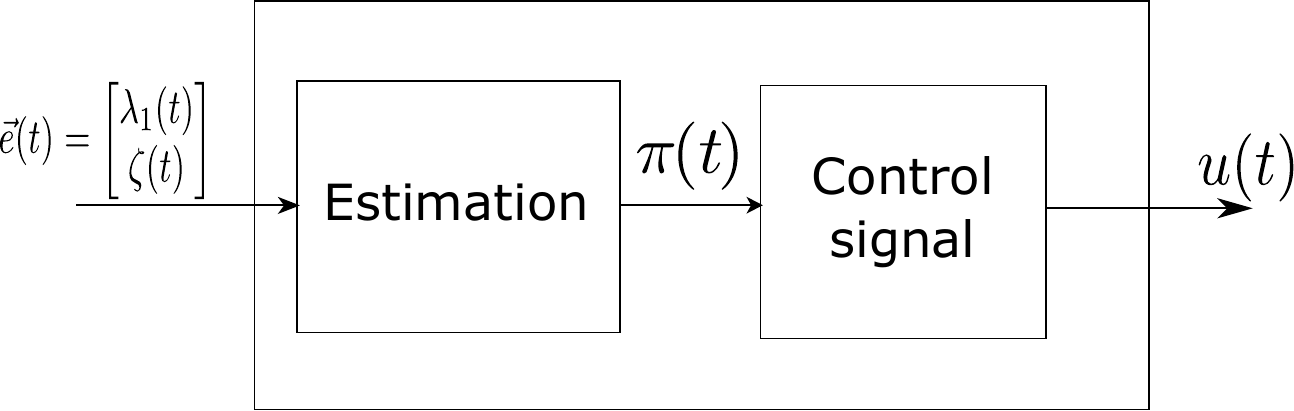}
\caption{Block diagram of the controller}
\label{fig:ctrl}
\end{figure}

Thus \refe{estimator2} and \refe{price} form the controller, which is illustrated in \reff{fig:ctrl}. The controller is implemented in two steps: in the first it estimates the average VOT based on $\lambda_1(t)$ and $\zeta(t)$, and in the second it calculates the dynamic price.
%\vspace*{-15pt}

\section{Analytical properties of the closed-loop control system with constant demand and vot} \label{properties}

With the definition of $\zeta(t)$ in \refe{zeta}, the feedback-based estimator in \refe{estimator2} and the dynamic price in \refe{price}, the point queue models in \refe{newpqmodel} and the lane choice model in \refe{newchoice-q3} form a closed-loop control system. There are six unknown variables in the system: $\lambda_1(t)$, $\lambda_2(t)$, $\zeta(t)$, $w(t)$, $\pi(t)$ and $u(t)$. In this section, we examine the analytical properties of the closed-loop control system with constant demand levels and a constant average VOT. Different from Section \ref{solution_control}, here we assume we don't know the true average VOT of SOVs.

\subsection{Equilibrium state}

Further by substituting \refe{price} into \refe{newchoice-q3}, we obtain the following system:
\begin{subequations}
\label{new-control-system}
\begin{align}
\der{}{t} \lambda_1(t)&=\max\{-\zeta(t), -\frac{\lambda_1(t)}\epsilon\},\label{la1-dynamics}\\
\der{}{t} \lambda_2(t)&=\max\{q_1+q_2-C_1-C_2+\zeta(t), -\frac{\lambda_2(t)}\epsilon\}, \label{la2-dynamics}\\
\zeta(t) &= C_1-q_1-\frac{q_2}{1+\exp(\alpha^*(u(t)-\pi^* w(t)))}, \label{q3_dynamics}\\
w(t)&=\frac{\lambda_2(t)}{C_2}-\frac{\lambda_1(t)}{C_1}, \label{w_dynamics}\\ 
\der{}{t} \pi(t) &=  K_1 \lambda_1(t) - K_2 \zeta(t), \label{pi_dynamics}\\
u(t)&=\pi(t)w(t)+ \frac{\ln\frac{q_1+q_2-C_1}{C_1-q_1}}{\alpha^*}.\label{u_dynamics}
\end{align}
\end{subequations}

We define the equilibrium state as when $\dot{\lambda}_1(t)=0$ and $\dot{\pi}(t)=0$, which leads to $\lambda_1(t)=\lambda_1$ and $\pi(t)=\pi^*$. From \refe{la1-dynamics} and \refe{pi_dynamics} we can see that, $\lambda_1=0$, and $\zeta(t)=\zeta=0$. From \refe{la2-dynamics} we find that, in the equilibrium state, $\der{}{t} \lambda_2(t)=q_1+q_2-C_1-C_2$. At a very large time $t$, the equilibrium queue size on the GP lanes is $\lambda_2^*(t)\approx (q_1+q_2-C_1-C_2)t$.

\subsection{Stability of the equilibrium state}\label{stability}

In this subsection, we analyze the local stability of the equilibrium state at a large time $t$, when $\lambda_1(t)$ and $\zeta(t)$ are both very small. In this case, substituting \refe{q3_dynamics} into \refe{pi_dynamics}, we obtain
\begin{align} \label{derive}
&-\frac {\dot {w(t)}}{\alpha^*w^2(t)} 
\ln (\frac{q_1+q_2-C_1+\zeta(t)}{C_1-q_1-\zeta(t)}\frac{C_1-q_1}{q_1+q_2-C_1} ) \notag\\
&+ \frac {1}{\alpha^*w(t)} 
 \frac {q_2} {(q_1+q_2-C_1+\zeta(t)) (C_1-q_1-\zeta(t))}  \der{}t \zeta(t) \notag \\& =K_1 \lambda_1(t)-K_2 \zeta(t).     
\end{align}

Since $\lambda_2(t)$ is very large, and $\zeta(t)$ is very small near the equilibrium state, from the above equation we have the following approximate dynamics for $\zeta(t)$:

\begin{equation}
\der{}t \zeta(t)\approx \beta t (K_1 \lambda_1(t)-K_2 \zeta(t)), \label{zeta-dynamics}   
\end{equation}
where $\beta=\frac{\alpha^* (q_1+q_2-C_1-C_2)(q_1+q_2-C_1)(C_1-q_1)}{C_2q_2}>0$.
Therefore, the system of \refe{la1-dynamics} and \refe{zeta-dynamics} approximates the dynamics of the original closed-loop control system, \refe{new-control-system}, at the equilibrium state subject to a small disturbance after a long time. Clearly the equilibrium state of the approximate system is at $(\lambda_1^*, \zeta^*)=(0,0)$. Note that \refe{zeta-dynamics} is a liner time-variant system; thus the approximate system is a switching linear time-variant system. 

\begin{theorem} \label{theorem}
The approximate system of \refe{la1-dynamics} and \refe{zeta-dynamics} is locally stable at the equilibrium state $(0,0)$ after a long time. Thus the closed-loop system, \refe{new-control-system}, is locally stable at its equilibrium state.
\end{theorem}

{\em Proof}: From \refe{la1-dynamics}, there are two phases for the queue dynamics on the HOT lanes.
\begin{enumerate}
    \item When $\zeta(t) \geq \frac{\lambda_1(t)}\epsilon$, $\lambda_1(t+\epsilon)=0$, and the queue on the HOT lanes vanishes. In this case, \refe{zeta-dynamics} can be simplified as
\begin{equation*}
\der{}t \zeta(t)\approx -\beta K_2 t \zeta(t).
\end{equation*}
The solution of the above equation is
\begin{equation}\label{solution1}  
\zeta(t) \approx \zeta(0) e^{-\frac 12 \beta K_2 t^2},  
\end{equation}

which converges to the equilibrium value $0$ in a Gaussian manner, much faster than exponentially.

\item When $\zeta(t) < \frac{\lambda_1(t)}\epsilon$, $\lambda_1(t)>0$, and the queue size on the HOT lanes is positive. In this case, \refe{la1-dynamics} can be written as
\begin{equation*}
\der{}t \lambda_1(t)=-\zeta(t).
\end{equation*}
At a very large time $t$, $\der{}t \zeta(t)$ is  finite, and \refe{zeta-dynamics} is equivalent to
\begin{equation*}
K_1 \lambda_1(t)-K_2 \zeta(t) \approx \frac1 {\beta t}\der{}t \zeta(t) \to 0.
\end{equation*}
Thus 
\begin{equation}\label{solution2}    
\zeta(t)\approx\frac {K_1}{K_2} \lambda_1(t)\approx \frac {K_1}{K_2} \lambda_1(0) e^{-\frac {K_1}{K_2} t}.
\end{equation}
which converges to the equilibrium state exponentially.

\end{enumerate}
In both cases, the approximate system is stable. \eop

We conclude that there are two convergence patterns of the approximate model after a long time:

Pattern 1: When $\lambda_1(t)=0$, $\zeta(t)$ converges in a Gaussian manner. 

Pattern 2: Both $\lambda_1(t)$ and $\zeta(t)$ converge exponentially, and $\frac{\lambda_1(t)}{\zeta(t)}=\frac{K_2}{K_1}$.

\section{Numerical examples}\label{simulation}
In this section, we provide numerical results of the methods above. The study site is a freeway segment with lane drop downstream of the GP lanes (see \reff{fig:traffic system}), and the capacity for one HOT and one GP lane is 30 veh/min. The downstream is not congested initially ($\lambda_1(0)=\lambda_2(0)=0$). The study period is 20 minutes, and the time-step size is $1/60$ min. For simplicity, We assume the true average VOT is $\$ 0.5$/min ($\pi^*= \$0.5$/min), and $\alpha^*=1$. Our initial guess of average VOT is $\$ 0.25$/min.

\subsection{Comparison of three controllers with constant demand} \label{constant}
We first consider the constant demand. The demand of HOVs is constant at $q_1(t)=10$ veh/min, and the demand of SOVs is constant at $q_2(t)=60$ veh/min. Then $q_3^*(t)=20$ veh/min. 

For the controller, \refe{estimator2}, we set $K_1= 0.1 ~ \mathrm{\$}/\mathrm{min}^{2}$ and $K_2=0.1$ \$/min.

\begin{figure}[h]
\centering
\subfloat[Queue sizes]{
  \includegraphics[width=0.48\textwidth]{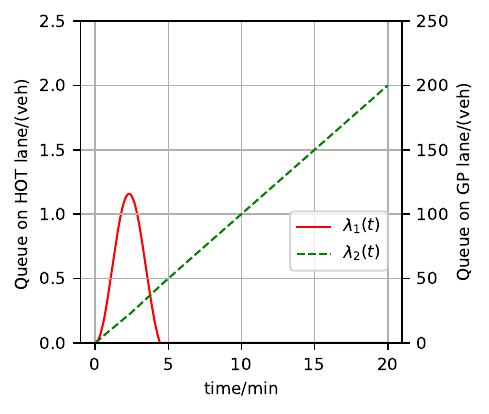}
  \label{fig: Simu1-q}}
  \hfill
\subfloat[Throughputs]{
  \includegraphics[width=0.48\textwidth]{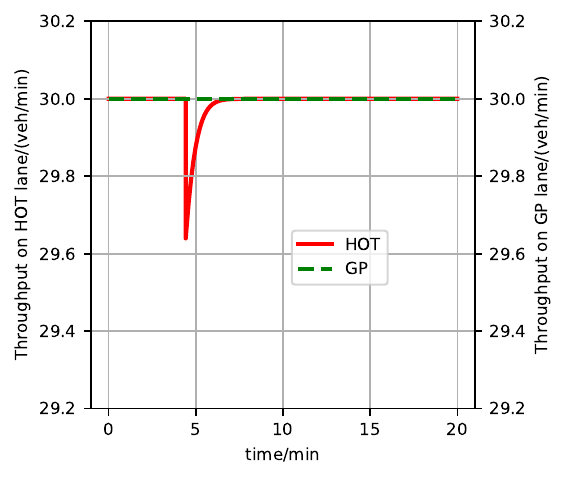}
  \label{fig: Simu1-throughput}}
\\
\subfloat[Estimated average VOT]{
  \includegraphics[width=0.48\textwidth]{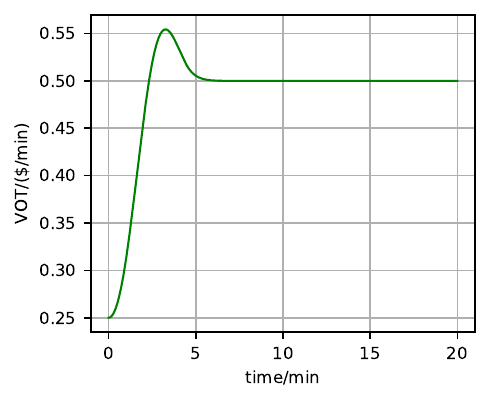}
  \label{fig: Simu1-vot}}
  \hfill
\subfloat[ Price]{
  \includegraphics[width=0.48\textwidth]{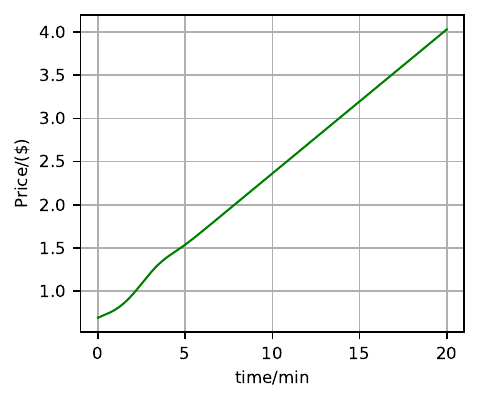}
  \label{fig: Simu1-u}}
\caption{Numerical results of our method (\refe{estimator2} and \refe{price}) with $K_1=0.1 ~ \mathrm{\$}/\mathrm{min}^{2}$ and $K_2=0.1$ \$/min}
\label{fig: Simu1}
\end{figure}

From Figure \ref{fig: Simu1-q}, we can observe some fluctuations in queue size at the beginning because the estimated average VOT is lower than the true value. After 2.5 minutes, the queue size on the HOT lanes starts to decrease and keeps at zero. Meanwhile, the queue on the GP lanes first increases at a rate lower than 10 veh/min, and at a higher rate between 2.5 and 7 minutes. After 7 minutes, it increases linearly with time. The HOT lanes are underutilized initially, but later the throughput is at capacity, as shown in Figure \ref{fig: Simu1-throughput}. The average throughput of the HOT lanes is 29.96 veh/min, and the throughput of the GP lanes is 30 veh/min since the GP lanes are always congested. Since there is no residual queue and no residual capacity on the HOT lanes, the control system can reach the optimal state after some time. In Figure \ref{fig: Simu1-vot}, we overestimate the average VOT initially, but after about 6 minutes, we can estimate the true average VOT ($\$0.5$/min). In Figure \ref{fig: Simu1-u}, the dynamic price increases linearly with time after the true average VOT is estimated, and the price is $\$ 4.024$ at 20 minutes. Then, we go back to the analytical results. From \refe{price_constant}, we can get $u(t)=\frac{1}{6} t+\ln 2$. At $t=20$ min, the corresponding price is $u(20)= \$ 4.026$. The optimal throughput of the HOT lanes is 30 veh/min. Comparing the numerical results with the analytical results, we find that our method indeed drive the system to the optimal state as expected.

\begin{figure}[h]
\centering
\subfloat[Queue sizes]{
  \includegraphics[width=0.48\textwidth]{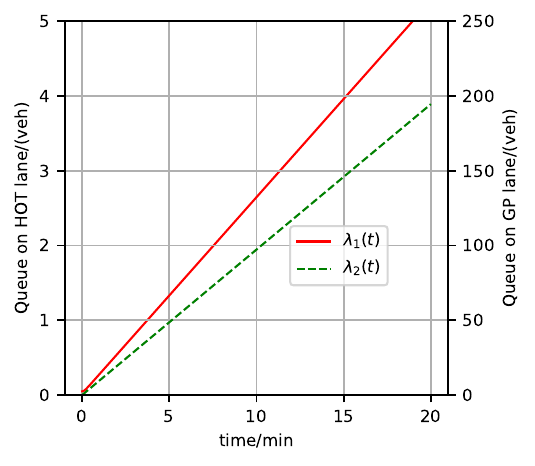}
  \label{fig:feedback_q}
  }
\hfill
\subfloat[Price]{
  \includegraphics[width=0.48\textwidth]{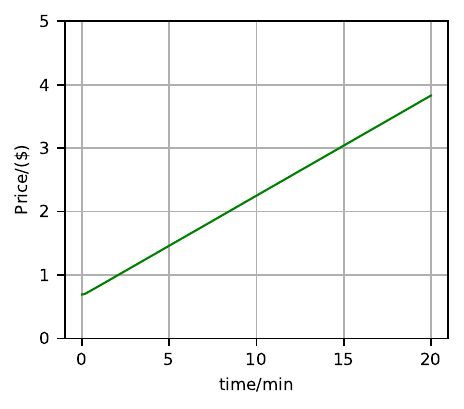}
  \label{fig:feedback_u}
  }
\caption{Numerical results of the feedback method in \cite{yin2009}}
\label{fig:feedback}
\end{figure}

For the feedback method \refe{feedback}, we set the initial price $u(0)=\ln 2$, and $K_I=0.01  (\mathrm{\$}\cdot\mathrm{min})$ for the controller. The numerical results are shown in \reff{fig:feedback}. Since the demand is always higher than the capacity of the HOT lanes, the price increases with time as shown in \reff{fig:feedback_u}, and the queue size in \reff{fig:feedback_q} increases as well. The zero-queue condition cannot be guaranteed by the I-controller. This verify our conclusion that the dynamic price determined by a single I-controller cannot achieve the two operation objectives for the HOT lanes at the same time. 

%\vspace{-20pt}

For the self-learning method \refe{self}, we set the true value to be $[\alpha_1; \alpha_2;\gamma]=[0.5;1;0]$, and the initial guess is $[\alpha_1(0); \alpha_2(0);\gamma(0)]=[0.25;1;0.1]$. And the variance of measurement noise is 0.09. 

\clearpage
\begin{figure}[h]
\centering
\subfloat[Queue sizes]{
  \includegraphics[width=0.48\textwidth]{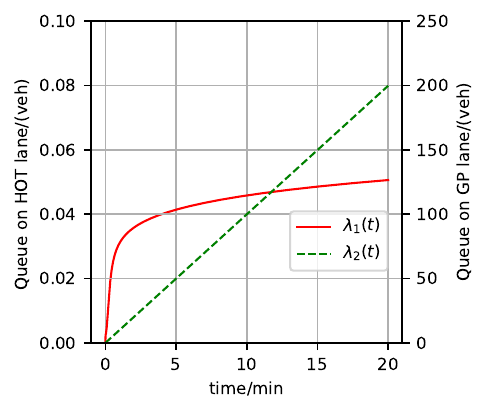}
  \label{fig: self-q}}
  \hfill
\subfloat[Throughputs]{
  \includegraphics[width=0.48\textwidth]{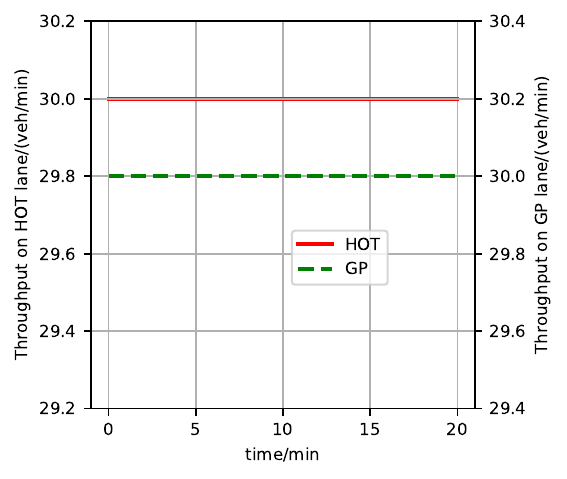}
  \label{fig: self-throughput}}
\\
\subfloat[ Estimated average VOT]{
  \includegraphics[width=0.48\textwidth]{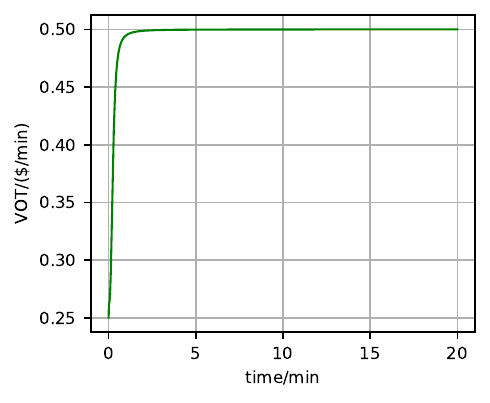}
  \label{fig: self-vot}}
  \hfill
\subfloat[ Price]{
  \includegraphics[width=0.48\textwidth]{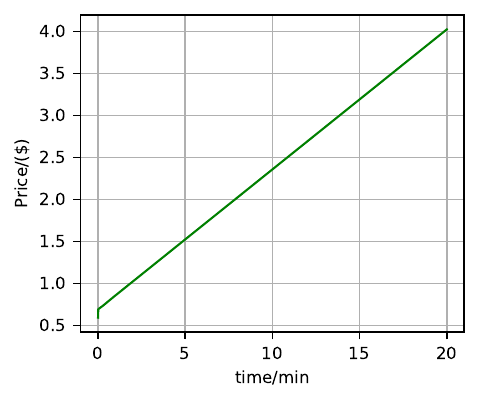}
  \label{fig: self-u}}
\caption{Numerical results of the self-learning method in \cite{yin2009}}
\label{fig: self_learning}
\end{figure}
\clearpage

Since the demand of paying SOVs is higher than the optimal value, queue size increases on the HOT lanes, as shown in \reff{fig: self-q}. So, the optimal state cannot be guaranteed. Since the demand of paying SOVs is not less than 20 veh/min, the throughput of the HOT lanes is always 30 veh/min. In \reff{fig: self-vot}, $\pi(t)$ rises from $\$0.25$/min, and converges to $\$0.5$/min. Different from \reff{fig: Simu1-vot}, there is no overshoot in the estimation process. In \reff{fig: self-u}, the dynamic price increases with time, and it is $\$ 4.024$ at 20 minutes. We also find that the length of the residual queue on the HOT is sensitive to the measurement noise and the initial guess of $\alpha_1$. The queue size will be longer if there is a larger measurement noise or a smaller initial guess of $\alpha_1$. 

Comparing with the system performance of those three methods, we conclude that our method is more effective.

\subsection{Robustness}\label{robustness}

In the previous section we considered a constant demand and VOT. That is not the case in reality: the demand varies with the time, the measurements are inaccurate subject to disturbances, and even the model parameters can change.  Robustness is the degree to which a system can function correctly in the presence of uncertainty. Providing robustness is one of the key uses for feedback control. In this subsection, we examine the robustness of the designed controller subject to disturbances in the demand pattern and VOT. We assume the demand of HOVs is Poisson with an average of 10 veh/min, and the demand of SOVs is Poisson with an average of 60 veh/min. For the controller, we keep $K_1=0.1 ~ \mathrm{\$}/\mathrm{min}^{2}$ and $K_2=0.1$ \$/min. 

\clearpage
\begin{figure}[h]
\centering
\subfloat[]{
  \includegraphics[width=0.48\textwidth]{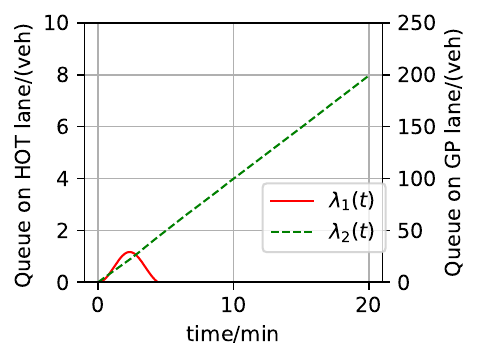}
  \label{fig: Simu2-q}}
  \hfill
\subfloat[]{
  \includegraphics[width=0.48\textwidth]{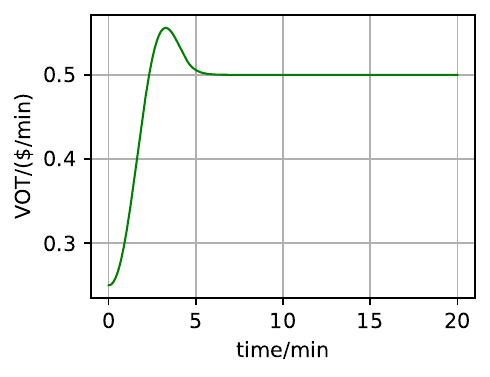}
  \label{fig: Simu2-vot}}
\\
\subfloat[]{
  \includegraphics[width=0.48\textwidth]{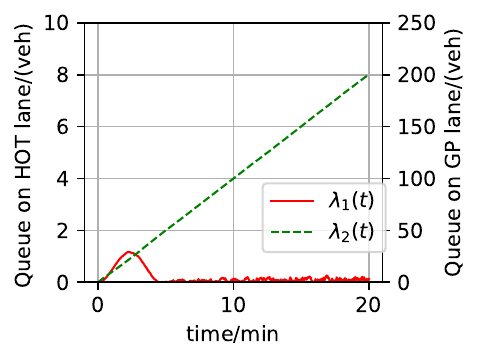}
  \label{fig: Simu2-q-2}}
  \hfill
\subfloat[]{
  \includegraphics[width=0.48\textwidth]{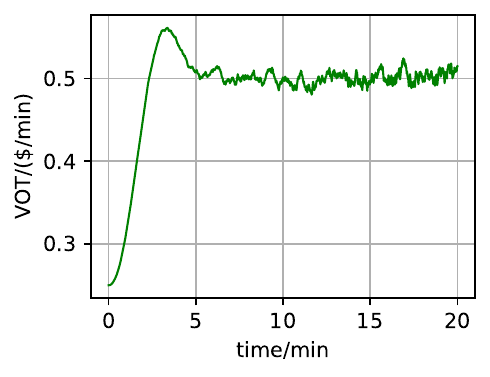}
  \label{fig: Simu2-vot-2}}
\caption{Numerical results of our method: (a) queue sizes with random demands; (b) estimated VOT with random demands; (c) queue sizes with random demands and other parameters or variables in the lane choice model; (d) estimated VOT with random demands and other parameters or variables in the lane choice model}
\label{fig: Simu2}
\end{figure}
\clearpage

With the random traffic demands, we can still estimate the average VOT, as shown in Figure \ref{fig: Simu2-vot}. Same as the results in Figure \ref{fig: Simu1-q}, no queue exists on the HOT lanes after the system reaches the optimal state (see Figure \ref{fig: Simu2-q}). 

We further add the randomness to other parameters or variables in the lane choice model in \refe{choice}, where $\eta \in [-0.1,0.1]$ follows a uniform distribution. Different from Figure \ref{fig: Simu2-q-2}, there is a residual queue on the HOT lanes. However, the queue size is close to zero. Thus the equilibrium state is Lyapunov stable. In Figure \ref{fig: Simu2-vot-2}, the estimated VOT fluctuates around the average VOT. In this sense, we can conclude that the controller is robust with respect to random variations in the demand patterns and other parameters or variables in the lane choice model.

\subsection{Stability of the system}

In this section, we provide numerical results of the original and the approximate model in Section \ref{properties}.

\subsubsection{The original model}
In this subsection, we numerically solve \refe{new-control-system} with different parameters, and subject to different initial perturbation to the equilibrium state. 

For the first pattern, we set $K_1=0.1 ~  \mathrm{\$}/\mathrm{min}^{2}$, $K_2=0.1$ \$/min, $\lambda_1(0)=1$ veh, and $\pi(0)=\$0.25$ /min. 

\begin{figure}[h]
\centering
\subfloat[]{
  \includegraphics[width=0.48\textwidth]{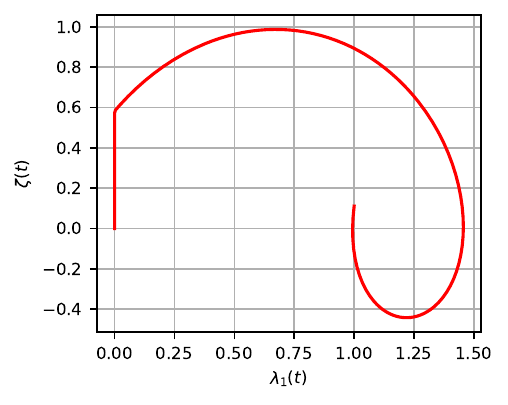}
  \label{phase1}
  }
\hfill
\subfloat[]{
  \includegraphics[width=0.48\textwidth]{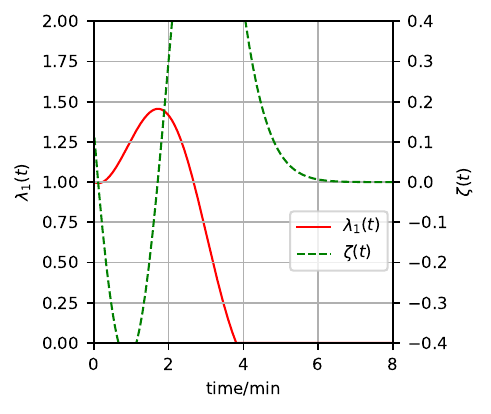}
  \label{timeseries1}
  }
\caption{Numerical results of the original model (convergence pattern 1 in \ref{theorem}).}
\label{fig:exact_1}
\end{figure}
\reff{phase1} is the phase diagram, the horizontal axis represents $\lambda_1(t)$, and the vertical axis represents $\zeta(t)$. The staring point is (1,0.11). Initially, $\zeta(t)$ decreases till reaches -0.44 veh/min, and the maximum $\lambda_1(t)$ is 1.46 veh. It is obvious that $\lambda_1(t)$ reaches 0 earlier than $\zeta(t)$. \reff{timeseries1} shows how $\lambda_1(t)$ and $\zeta(t)$ change with the time. The horizontal axis is time, the left vertical axis represents $\lambda_1(t)$ and the right vertical axis represents $\zeta(t)$. It is clear that $\lambda_1(t)$ reaches 0 at around 4 minutes, and after that $\zeta(t)$ converges to 0 in a Gaussian manner (Theorem \ref{theorem}).

For the second pattern, we set $K_1=0.1 ~ \mathrm{\$}/\mathrm{min}^{2}$, $K_2=0.2$ \$/min, $\lambda_1(0)=1$ veh and $\pi(0)=\$0.25$ /min. 

\begin{figure}[h]
\centering
\subfloat[]{
  \includegraphics[width=0.48\textwidth]{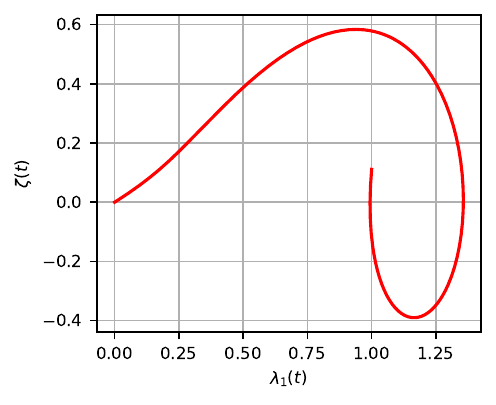}
  \label{phase2}}
  \hfill
\subfloat[]{
  \includegraphics[width=0.48\textwidth]{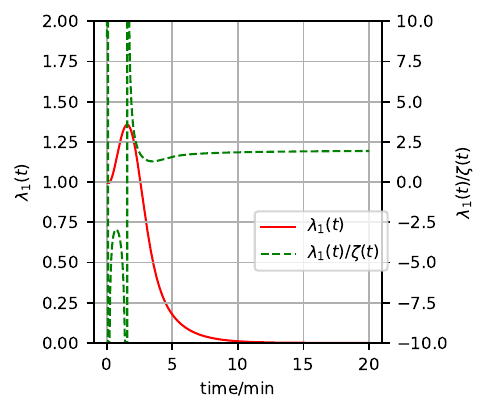}
  \label{timeseries2}}
\caption{Numerical results of the original model (convergence pattern 2 in \ref{theorem})}
\label{fig:exact_2}
\end{figure}

In \reff{phase2}, the starting point is (1,0.11). The minimum value of $\zeta(t)$ is -0.39 veh/min, and the maximum value of $\lambda_1(t)$ is 1.36 veh. Both $\lambda_1(t)$ and $\zeta(t)$ converges to 0 exponentially after a long time. In \reff{timeseries2}, the horizontal axis is time, and the left vertical axis represents $\lambda_1(t)$. Different from \reff{timeseries1}, the right vertical axis represents the ratio between $\lambda_1(t)$ and $\zeta(t)$. $\lambda_1(t)$ converges exponentially staring from 2 minutes, and reaches 0 at around 20 minutes. $\lambda_1(t)/\zeta(t)$ converges to 2, which equals $K_2/K_1$ (Theorem \ref{theorem}).

We further numerically check the convergence pattern. With $\lambda_1(0)=1$ veh and $\pi(0)= \$0.25 $ /min, when we set $K_1=0.1 ~ \mathrm{\$}/\mathrm{min}^{2}$, the two phases switch around $K_2=0.14$ \$/min. From \reff{fig:exact_1} and \ref{fig:exact_2}, we observe that a larger $K_2$ leads to congested, exponential convergence; a smaller $K_2$ leads to uncongested, Gaussian convergence (Theorem \ref{theorem}).

When the demands are stochastic, we can observe similar convergence patterns to \reff{fig:exact_1} and \ref{fig:exact_2}. 

\subsubsection{The approximate model}\label{switch}

In this subsection, we numerically solve the approximate model of \refe{la1-dynamics} and \refe{zeta-dynamics}.

For the first pattern, we set $K_1=0.1 ~ \mathrm{\$}/\mathrm{min}^{2}$, $K_2=0.1$ \$/min, $\lambda_1(0)=1$ veh and $\zeta(0)= 0.11$ veh/min. 

\begin{figure}[h]
\centering
\subfloat[]{
  \includegraphics[width=0.48\textwidth]{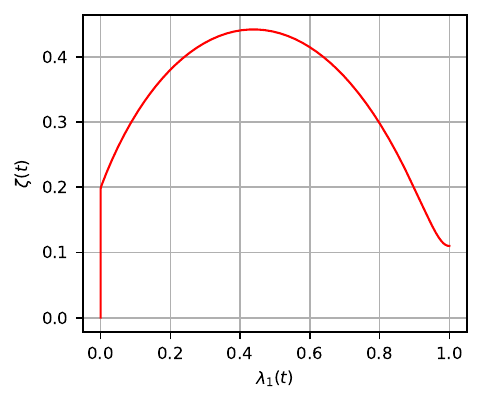}
  \label{phase3}
  }
\subfloat[]{
  \includegraphics[width=0.48\textwidth]{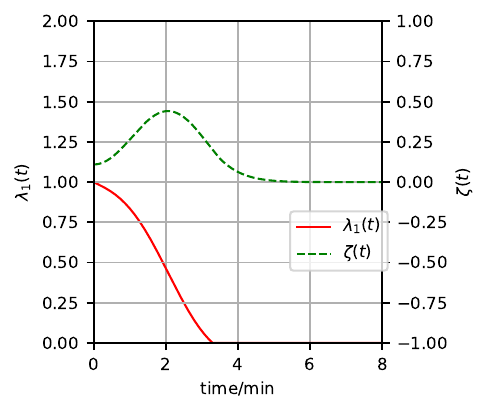}
  \label{timeseries3}
  }
\caption{Numerical results of the approximate model (convergence pattern 1 in \ref{theorem})}
\label{fig:switch_1}
\end{figure}
In \reff{phase3}, the trajectory starts from (1,0.11), and ends at (0,0). The maximum value of $\zeta(t)$ is 0.44 veh/min. The convergence pattern is different from \reff{phase1} initially, because we drop the first term on the left-hand side of \refe{derive} when deriving \refe{zeta-dynamics}. After a long time, the convergence pattern is similar to \reff{phase1}. From \reff{timeseries3}, we can observe that $\lambda_1(t)$ decreases until it reaches 0, while $\zeta(t)$ increases to 0.44 and then converges to 0 in a Gaussian manner (Theorem \ref{theorem}).

For the second pattern, we set $K_1=0.1 ~ \mathrm{\$}/\mathrm{min}^{2}$, $K_2=0.2$ \$/min, $\lambda_1(0)=1$ veh and $\zeta(0)= 0.11$ veh/min. 

\begin{figure}[h]
\centering
\subfloat[]{
  \includegraphics[width=0.48\textwidth]{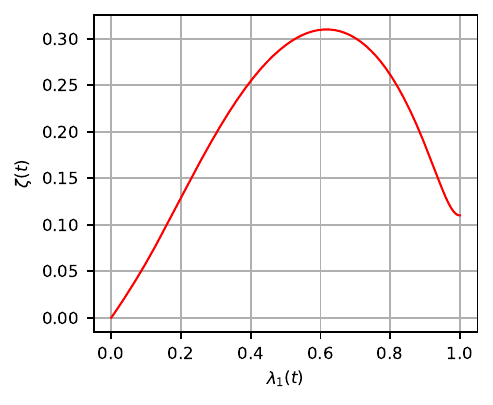}
  \label{phase4}
  }
\subfloat[]{
  \includegraphics[width=0.48\textwidth]{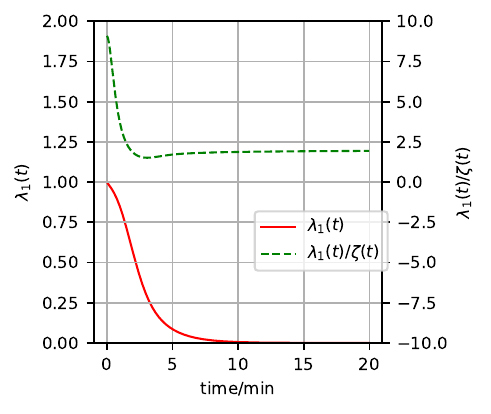}
  \label{timeseries4}
  }
\caption{Numerical results of the approximate model (convergence pattern 2 in \ref{theorem})}
\label{fig:switch_2}
\end{figure}
In \reff{phase4}, the convergence pattern is similar to \reff{phase2} after a long time. The starting point is (1,0.11), and ends at (0,0). The maximum value of $\zeta(t)$ is 0.31 veh/min. In \reff{timeseries4}, $\lambda_1(t)$ converges to 0 exponentially (Theorem \ref{theorem}); and $\lambda_1(t)/\zeta(t)$ converges to 2 after a long time, which is the same as \reff{timeseries2}. 

We also numerically check the convergence pattern. With $\lambda_1(0)=1$ veh and $\zeta(0)= 0.11$ veh/min, when we set $K_1=0.1 ~ \mathrm{\$}/\mathrm{min}^{2}$, the two phases switch around $K_2=0.14$ \$/min. From \reff{fig:switch_1} and \ref{fig:switch_2}, we observe that a larger $K_2$ leads to congested, exponential convergence; a smaller $K_2$ leads to uncongested, Gaussian convergence (Theorem \ref{theorem}).

\subsection{The scale parameter}

In this section, we discuss the case when the operators don't know the true value of the scale parameter. Instead, they make a guess of the scale parameter, denoted as $\alpha$, when determining the dynamic price. Then, the price is set as
\begin{equation}\label{newprice}
u(t)=\pi(t)w(t)+\frac{\ln\frac{q_1+q_2-C_1}{C_1-q_1}}{\alpha}.
\end{equation}

The setup is the same as Section \ref{constant}, and $\alpha=1.2$.

\clearpage
\begin{figure}[h]
\centering
\subfloat[Queue sizes]{
  \includegraphics[width=0.48\textwidth]{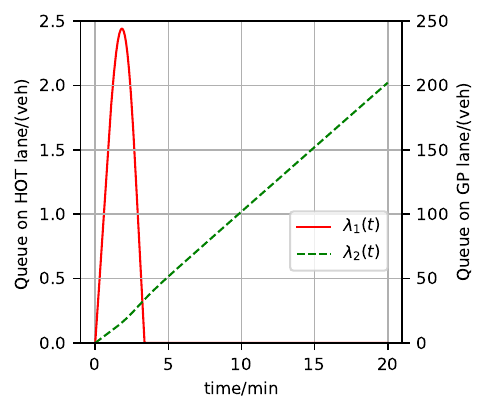}
  \label{fig: Simu3-q}}
  \hfill
\subfloat[Throughputs]{
  \includegraphics[width=0.48\textwidth]{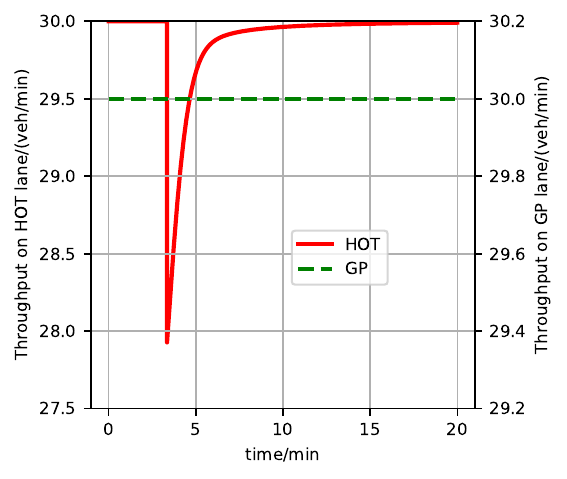}
  \label{fig: Simu3-throughput}}
\\
\subfloat[ Estimated average VOT]{
  \includegraphics[width=0.48\textwidth]{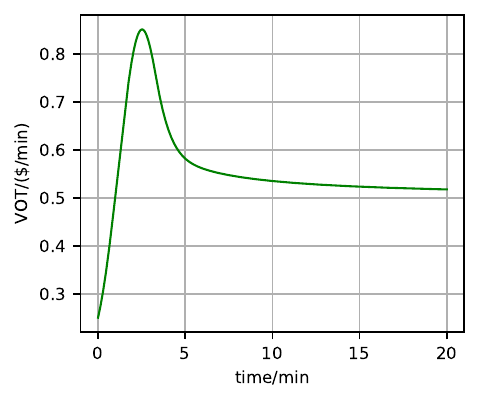}
  \label{fig: Simu3-vot}}
  \hfill
\subfloat[ Price]{
  \includegraphics[width=0.48\textwidth]{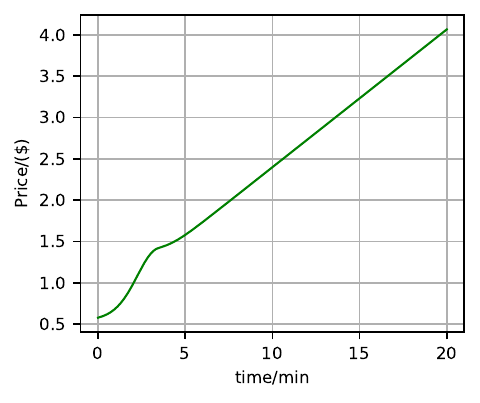}
  \label{fig: Simu3-u}}
\caption{Numerical results of our method (\refe{estimator2} and \refe{newprice}) with $K_1=0.1 ~ \mathrm{\$}/\mathrm{min}^{2}$ and $K_2=0.1$ \$/min}
\label{fig: Simu3}
\end{figure}
\clearpage

In \reff{fig: Simu3-q}, no queue exists on the HOT lanes after some time, which is the same as \reff{fig: Simu1-q}. The throughputs on both lanes are 30 vph eventually (see \reff{fig: Simu3-throughput}). So, the price in \refe{newprice} drives the system to the optimal state. In \reff{fig: Simu3-vot}, the estimated average VOT is approaching the true value. The price is $\$ 4.061$ at 20 minutes in \reff{fig: Simu3-u}, which is slightly higher than the one in \reff{fig: Simu1-u}. 

\begin{figure}[h]
\centering
\includegraphics{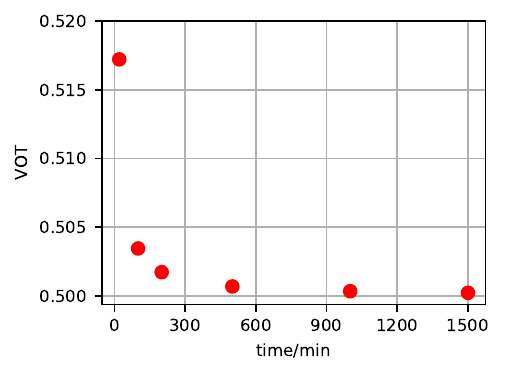}
\caption{The estimated VOT}
\label{fig:VOT}
\end{figure}

After a long time, $\pi(t)$ converges to $\pi^*$ (see \reff{fig:VOT}). The result is consistent with the statement that the VOT is not affected by the scale parameter \cite{train2009discrete}.
\vspace*{-10pt}

\section{Conclusion}\label{conclusion}

In this study, we provide a new control theoretic approach to solve the simultaneous estimation and control problem for a traffic system with HOT lanes. We define a new variable called the residual capacity for the HOT lanes. We apply two point queue models to capture traffic dynamics on the HOT and GP lanes, and capture the lane choice of SOVs with a logit model. The controller can estimate the average VOT and calculate the dynamic price simultaneously, which has received little attention in the field of traffic control. We also analytically prove that the closed-loop system is stable and guaranteed to converge to the optimal state. We provide numerical examples to show that our method is effective and robust with respect to randomness in demands and parameters or variables in the lane choice model. We find the system converges to the equilibrium state in the patterns predicted by Theorem \ref{theorem}. At the same time, we show that the two methods in \cite{yin2009} are either unstable or cannot guarantee the convergence to the optimal state, and our controller is more efficient and leads to a better system performance.
Before designing a specific controller for the system, we present a simple analytical example to show that when the HOT lanes are underutilized while the whole system is congested, the price for the HOT lanes should increase with time. This gives us an guideline for the controller design. Then, we design a simple I-controller to estimate the average VOT of SOVs. We also provide a novel analytical proof of the stability, since such a theoretic study on the stability was not available in previous researches on the HOT lanes. At last, we numerically show that the scale parameter does not affect the estimation of VOT and the optimal state after a long time.

In summary, this study makes three fundamental contributions: (i) to present a simpler formulation of the point queue model based on the new concept of residual capacity, (ii) to propose a simple feedback control theoretic approach to estimate the average value of time and calculate the dynamic price, and (iii) to analytically and numerically prove that the closed-loop system is stable and guaranteed to converge to the optimal state, in either Gaussian or exponential manners. The methodology and result are novel and relevant both theoretically and practically, and the estimator/controller as well as the analytical method can enable us for better understanding and design of effective and efficient dynamic pricing strategies for real-world systems in the future.

The following are some potential future research topics.
\begin{itemize}

\item
Queue detection problem is fundamental in traffic control. To measure the queue size on HOT and GP lanes, we need two sets of loop detectors on each lane group.  For each lane group, the queue size is the difference between the cumulative arrival and departure flows. Cao et al. proposed a detection method based on the shockwave theory \cite{cao2015real}. We will be interested in developing new methods to measure queues on freeways.  

\item
We will be interested in the estimation and control problem in a more complex network. We will consider the effect of lane changing \cite{jin2013multi} and capacity drop \cite{jin2015kinematic}. 
%We will also consider other traffic flow models to describe the dynamics.

%\item
%In this study, we capture the lane choice by a logit model with the average VOT. In the future, we want to introduce a new choice model called \enquote{vehicle-based} user equilibrium. The basic idea is as follows: if a single SOV chooses a lane (HOT or GP), then the cost of this lane is less than or equal to that of the other lane (GP or HOT) which would be experienced by the same vehicle. With the \enquote{vehicle-based} UE, we can consider the heterogeneous VOTs for different SOVs.
\item 
In a HOT lane system, the choice of SOVs is quite complex \cite{janson2018choice,burris2018unrevealed}. We will be interested in developing new models to better capture the choice behavior for a traffic system with HOT lanes. 

\item
Another research topic would be a simultaneous departure time and lane choice model for SOVs \cite{boyles2015incorporating}. For SOVs, their travel costs include a free-flow travel time, a queuing time, a schedule delay and a dynamic price. We want to design pricing schemes considering departure time user equilibrium.

\item
Conceptually, an underlying assumption is that the HOT lanes are Pareto improving compared with the HOV lanes; i.e., both paying and non-paying SOVs are better off, and the HOVs are not negatively affected. Such an assumption is only valid when the HOV lanes are underutilized and the GP lanes are congested. We will be interested in examining the situations when these conditions are violated.
\end{itemize}

\section*{Acknowledgment}

This research is partially sponsored by  the ITS-Irvine Mobility Research Program (SB1). The views and results are the authors' alone.

\end{document}